\documentclass[preprint,aps,showpacs,superscriptaddress]{revtex4}
\usepackage{graphicx}
\usepackage{dcolumn}
\usepackage{amsfonts, amssymb}
\begin{document}

\title{One- versus two-pole $\bar{K}N - \pi \Sigma$
potential: $K^- d$ scattering length}

\author{N.V. Shevchenko}
\affiliation{Nuclear Physics Institute, 25068 \v{R}e\v{z}, Czech Republic}

\date{\today}

\begin{abstract}
We investigated the dependence of the $K^- d$ scattering length
on models of $\bar{K}N$ interaction with one or two poles
for $\Lambda(1405)$ resonance. The $\bar{K}NN - \pi \Sigma N$ system is
described by coupled-channel Faddeev equations in AGS form. Our new two-body
$\bar{K}N - \pi \Sigma$ potentials reproduce all existing experimental data on
$K^- p$ scattering and kaonic hydrogen
atom characteristics. New models of $\Sigma N - \Lambda N$ interaction
were also constructed. Comparison with several approximations, usually
used for scattering length calculations, was performed.
\end{abstract}

%13.75.Jz: kaon-baryon interactions
%11.80.Gw: Multichannel scattering
%36.10.Gv: Mesonic atoms and molecules, hyperonic atoms and molecules
\maketitle

\section{Introduction}
\label{Introduction_sect}

Investigation of $K^- d$ system can shed more light on the $\bar{K}N$ interaction,
necessary for study of antikaonic nuclear clusters, which attracted
large interest recently~\cite{ECT2009}. The interaction is not very well
known, in particular, there are debates about the nature of the $\Lambda(1405)$
resonance. The question is whether it is a single resonance in $\pi \Sigma$ and a
quasi-bound state in $\bar{K}N$ channel or the bump, which is usually understood
as $\Lambda(1405)$ resonance, is an effect of two poles. 
The advantage of the $K^-d$ system is the possibility of proper description
of its dynamics using Faddeev equations~\cite{Faddeev}.

Recently we constructed coupled-channel $\bar{K}N- \pi \Sigma$ potentials
in one- and two-pole form~\cite{ourPRC_isobreak}, which reproduce all existing
experimental data on $K^- p$ scattering and $K^- p$ atom characteristics equally
well, therefore it is not possible to give preference to any of the versions.
A possible way to clarify the question concerning the nature of the $\Lambda(1405)$
resonance is to perform few- or many-body calculations using one- and two-pole
$\bar{K}N-\pi \Sigma$ potentials as an input.
Having in mind SIDDHARTA experiment~\cite{SIDDHARTA}, measuring characteristics
of kaonic deuterium atom, we calculated $K^- d$ scattering length $a_{K^- d}$
and investigated the dependence of the results on the models
of $\bar{K}N$ interaction with newly obtained parameters.
The scattering length gives possibility to calculate kaonic deuterium
level shift and width.  Comparison of the theoretical results with
experimental ones could allow to choose between the two $\bar{K}N - \pi \Sigma$
interaction versions.

Dependence of $a_{K^- d}$ on other two-body interactions, necessary for the
description of the $\bar{K}NN - \pi \Sigma N$ system, was also investigated:
we used several models of $NN$ (with and without short range repulsion)
and $\Sigma N (- \Lambda N)$ interactions.
In addition to the full coupled-channel calculation we performed checks of
commonly used approximations for $K^- d$ scattering. In particular, we solved
one-channel Faddeev equations using exact optical and simple complex $\bar{K}N$
potentials approximating the $\bar{K}N - \pi \Sigma$ models of interaction. We
also checked the ``Fixed center approximation to Faddeev equations'' formula.

The formalism used for the coupled-channel $K^- d$ scattering length calculation is
described in the next section. Section~\ref{two-bodyInp.sect} is devoted to
the two-body input: the description of the one- and two-pole
$\bar{K}N - \pi \Sigma$ potentials with newly obtained parameters
in the first subsection is supplemented with additional arguments for equivalence
of the two versions. The following subsections of Section~\ref{two-bodyInp.sect}
are devoted to $NN$ and $\Sigma N - \Lambda N$ potentials.
Section~\ref{approximate.sect} contains information
about approximate methods, usually used in $K^- d$ scattering length calculations.
The full and approximate results are shown and discussed in
Section~\ref{Results.sect}, while Section~\ref{conclusions.sect} concludes the paper.

\section{Coupled-channel AGS equations for $\bar{K}NN - \pi \Sigma N$ system}
\label{AGS_eq.sect}

As in~\cite{our_KNN_PRL,our_KNN_PRC} we directly include $\pi \Sigma N$
channel  into original three-body Faddeev equations in the Alt-Grassberger-Sandhas (AGS)
form~\cite{AGS}, which leads to the coupled-channel equations:
\begin{equation}
\label{U_coupled}
U_{ij}^{\alpha \beta} = {\delta}_{\alpha \beta} \,(1-\delta_{ij})
\, \left(G_0^{\alpha} \right)^{-1} + \sum_{k, \gamma=1}^3 (1-\delta_{ik})
\, T_k^{\alpha \gamma} \, G_0^{\gamma} \, U_{kj}^{\gamma \beta},
\end{equation}
where ``particle channel'' indices $\alpha,\beta=1,2,3$ are
introduced in addition to the usual Faddeev partition indices $i,j,k=1,2,3$,
see Table I of~\cite{our_KNN_PRC}. The equations define unknown operators 
$U_{ij}^{\alpha \beta}$, describing the elastic and re-arrangement processes
$j^{\beta} + (k^{\beta} i^{\beta}) \to i^{\alpha} + (j^{\alpha} k^{\alpha})$.
The free Green's function is diagonal in channel indices:
$G_0^{\alpha \beta} = \delta_{\alpha \beta} \, G_0^{\alpha}$.
The inputs for the system of equations~(\ref{U_coupled}) are two-body $T$-matrices,
embedded into three-body space: $T_i^{\alpha \beta}$ describes the interaction between
the particles $j$ and $k$ ($i \ne j \ne k$) in channels $\alpha,\beta$.
Like in~\cite{our_KNN_PRC}, here we have $T_i^{NN}$, $T_i^{\pi N}$ and
$T_i^{\Sigma N}$, which are usual one-channel two-body $T$-matrices in three-body
space, describing
$NN$, $\pi N$, and $\Sigma N$ interactions, respectively.
The $T_i^{KK}$, $T_i^{\pi \pi}$, $T_i^{\pi K}$, and $T_i^{K \pi}$ are
elements of the coupled-channel $T$-matrix for the $\bar{K}N - \pi \Sigma$
system.

In contrast to the calculation of the quasi-bound $K^- pp$ state~\cite{our_KNN_PRC},
where one-term isospin ($I$) dependent separable potentials were used,
now we write AGS equations for N-term isospin dependent separable potentials
\begin{equation}
\label{V_Nterm}
 V_{i,I}^{\alpha \beta} = \sum_{m=1}^{N_i^{\alpha}}
 \lambda_{i(m),I}^{\alpha \beta} \,
 |g_{i(m),I}^{\alpha} \rangle  \langle g_{i(m),I}^{\beta} |,
\end{equation}
which lead to separable $T$-matrices
\begin{equation}
\label{T_Nterm}
 T_{i,I}^{\alpha \beta} = \sum_{m,n = 1}^{N_i^{\alpha}}
 |g_{i(m),I}^{\alpha} \rangle
\tau_{i(mn),I}^{\alpha \beta} \langle g_{i(n),I}^{\beta} | \,.
\end{equation}
Here $N_i^{\alpha}$ is a number of terms of the separable potential, $\lambda$
is a strength constant, while $g$ is a form-factor. Bound state
wave function of the two-body subsystem, described by such a potential, has
the form
\begin{equation}
\label{psi}
 |\psi_{i,I}^{\alpha} \rangle = \sum_{m=1}^{N_i^{\alpha}}
 C_{i(m),I}^{\alpha} \, G_0^{\alpha}(z_{\,bnd}) \,
 |g_{i(m),I}^{\alpha} \rangle \,,
\end{equation}
where the coefficients $C_{i(m),I}^{\alpha}$ are constants and $z_{\,bnd}$ is a
binding energy. 
We used two slightly different versions of a two-term nucleon-nucleon separable
potential ($N_{i}^{\alpha} = 2$) in the $a_{K^- d}$ calculations. All other models of
interactions, one more $V^{NN}$ among them,  are one-term potentials with
$N_{i}^{\alpha} = 1$. 

The amplitude $f_{ij,I_i I_j}^{\alpha \beta}$ of the $\bar{K}(NN) \to \bar{K}(NN)$ reaction
with $NN$ in isospin-zero state in the initial and final states is defined by the
following matrix element:
\begin{equation}
\label{ampl}
f_{11,00}^{11}(\vec{p}_1^{\,1},\vec{p '}_1^{\,1};z_{tot}) = - (2 \pi)^2 \,
\mu_1^1 \, \langle \vec{p}_{1}^{\,1}; \psi_{1,0}^1 | U_{11,00}^{11}(z_{tot}) |
\psi_{1,0}^1; \vec{p '}_{1}^{\,1} \rangle \,,
\end{equation}
where $\mu_i^{\alpha}$ is the three-body reduced mass defined by
\begin{equation}
\label{mu}
\mu_i^{\alpha} = \frac{m_i^{\alpha} (m_j^{\alpha} + m_k^{\alpha})}
{m_i^{\alpha} + m_j^{\alpha} + m_k^{\alpha}},
\qquad i\neq j\neq k,
\end{equation}
$\vec{p}_1^{\,1}$ and $\vec{p '}_1^1$ are initial and final relative momenta of antikaon
with respect to the $NN$ pair, correspondingly, while $z_{tot}$ is the total
energy of the three-body system. Deuteron wave function $\psi_{1,0}^1$ in Eq.(\ref{ampl})
is defined by Eq.(\ref{psi}), the transition operator $U_{ij,I_i I_j}^{\alpha \beta}$ has
 two additional isospin indices compared to Eq.(\ref{U_coupled}).
The scattering length is the amplitude at zero kinetic energy $z_{kin}^{\alpha}$:
\begin{equation}
\label{aKd}
a_{K^- d} =
f_{11,00}^{11}(\vec{p}_1^{\,1} \to 0,\vec{p '}_1^{\,1} \to 0;z_{tot} \to z_{th}^1)
\end{equation}
with $z_{th}^{\alpha} = \sum_{i=1}^3 m_i^{\alpha}$ being the
$\bar{K}NN$ ($\alpha = 1$) or $\pi \Sigma N$ ($\alpha = 2$) threshold energy
(the total energy is defined by $z_{tot} = z_{kin}^{\alpha} + z_{th}^{\alpha}$).

Taking into account forms of $T$-matrices~(\ref{T_Nterm}) and wave functions~(\ref{psi}),
introducing new operators
\begin{eqnarray}
\label{X_definition}
&{}&
X_{i(m)j, I_i I_j}^{\alpha \beta} \equiv  \langle
g_{i(m),I_i}^{\alpha} | G_0^{\alpha} \, U_{ij, I_i I_j}^{\alpha
\beta} | \psi_{j,I_j}^{\beta} \rangle \,, \\
\label{Z_definition}
&{}&
Z_{i(m)j(n), I_i I_j}^{\alpha \beta} \equiv
\delta_{\alpha \beta} \, Z_{i(m)j(n), I_i I_j}^{\alpha} =
\delta_{\alpha \beta} \, (1-\delta_{ij}) \,
\langle g_{i(m),I_i}^{\alpha} | G_0^{\alpha} | g_{j(n),I_j}^{\alpha}
\rangle \,,
\end{eqnarray}
and substituting them into the system~(\ref{U_coupled}),
we can write a system of equations for the new unknown operators
$X_{ij, I_i I_j}^{\alpha \beta}$:
\begin{eqnarray}
\label{full_oper_eq}
X_{i(l)j, I_i I_j}^{\alpha \beta} &=& \delta_{\alpha \beta} \,
\sum_{m=1}^{N_j^{\alpha}} C_{j(m)}^{\alpha} \,
Z_{i(l)j(m), I_i I_j}^{\alpha} +  \\
\nonumber
&+& \sum_{k,\gamma=1}^3 \sum_{m,n=1}^{N_k^{\alpha}} \sum_{I_k}
Z_{i(l)k(m), I_i I_k}^{\alpha} \, \tau_{k(mn), I_k}^{\alpha \gamma} \,
X_{k(n)j, I_k I_j}^{\gamma \beta} \,.
\end{eqnarray}
The number of equations in the system (\ref{full_oper_eq}) is defined by the number
of all form-factors $g$. Therefore, the system~(\ref{full_oper_eq})
with a two-term $NN$ and one-term other potentials consists of 20 equations.

Two identical nucleons, entering the first ($\bar{K}NN$) channel, require antisymmetrization
of the system of equations. Orbital momentum of all two-body interactions was set to zero.
The main $\bar{K}N - \pi \Sigma$ potential was constructed with orbital momentum $l=0$ 
since the interaction is dominated by the $s$-wave $\Lambda(1405)$ resonance. The
interaction of $\pi$-meson with the nucleon is mainly in $p$-wave, however, as it was shown
in~\cite{Kd_BFMS}, the addition of ``small two-body interactions'', $\pi N$ among
them, changes the resulting $a_{K^- d}$ very slightly (of the order of $1 \%$ or even less,
see Table XIII of the paper). On the other hand, $s$-wave $\pi N$ interaction is
even weaker, therefore, we omitted $\pi N$ interaction in our equations.
Information about $\Sigma N$ interaction is very poor, and there is no reason to
assume significant effect of higher partial waves. Finally, $NN$ interaction 
was also taken in $l=0$ state only since we do not see physical reasons for
sufficient effect of higher partial waves in the present calculation.

Antisymmetric $s$-wave deuteron wave function has zero isospin and spin
equal to one. Due to this $K^- d$ system, in contrast to $K^- pp$~\cite{our_KNN_PRL,our_KNN_PRC},
has total three-body spin (and total momentum) equal to one, while both $\bar{K}NN$
systems have total isospin $I = \frac{1}{2}$. Therefore, in the $K^- d$ case
antisymmetrization leads to the following new operators:
\begin{equation}
\label{X_symmetrical}
\begin{array}{ll}
X_{1(m),0}^{1,asm} = X_{1(m),0}^{1},  & \quad
 X_{2,I}^{1,asm} = X_{2,I}^{1} - X_{3,I}^{1}, \\
X_{1,\frac{1}{2}}^{2,asm} = X_{1,\frac{1}{2}}^{2} + X_{1,\frac{1}{2}}^{3},
& \quad
 X_{1,\frac{3}{2}}^{2,asm} = X_{1,\frac{3}{2}}^{2} - X_{1,\frac{3}{2}}^{3}, \\
X_{2,I}^{2,asm} = X_{2,I}^{2} - X_{3,I}^{3},
& \quad
 X_{3,I}^{2,asm} = X_{3,I}^{2} - X_{2,I}^{3} \,.
\end{array}
\end{equation}
It is necessary to note, that the $\bar{K}^0 nn$ state drops out from the
system of equations (\ref{full_oper_eq}) after the antisymmetrization
because the two neutrons are in isospin one state. Therefore the $\bar{K}^0 nn$
channel has another value of the three-body spin ($S=0$) than $K^- d$ ($S=1$)
or the neutrons does not satisfy Pauli principle.

Finally, $K^- d$ scattering length can be found from
\begin{equation}
a_{K^- d} = - (2 \pi)^2 \, \mu_1^1 \,
\sum_{m=1}^{2} \, C_{1(m),0}^{1} \,
X_{1(m),0}^{1,asm}(0, 0; z_{th}^1) \,.
\end{equation}
The operator system~(\ref{full_oper_eq}) written in momentum space turns into
a system of integral equations. In order to solve the inhomogeneous system we
transformed the integral equations into algebraic ones. It is known
(see. e.g.~\cite{VB}), that integral Faddeev equations have moving
logarithmic singularities in the kernels when scattering above a 
three-body threshold ($z_{kin} > 0$) is described. ``Usual'' (one-channel)
scattering length calculations
are free of the singularities. For the $\bar{K}NN - \pi \Sigma N$ system,
however, where $z_{th}^2 < z_{th}^1$, the permanently opened $\pi \Sigma N$
channel causes appearance of logarithmic singularities in $K^- d$ scattering
length calculations.
In the numerical procedure we handle them using the method suggested
in~{\cite{log_sing}}. The main idea of the method consists in interpolating
the unknown solutions (in the interval containing the singular points) by
certain polynomials and subsequent analytic integration of the singular part
of the kernels.

\section{Two-body input}
\label{two-bodyInp.sect}

The separable $\bar{K}N - \pi \Sigma$, $NN$, and $\Sigma N$
potentials~(\ref{V_Nterm}) in momentum representation have the form:
\begin{equation}
\label{Vseprb}
 V_{I}^{\bar{\alpha} \bar{\beta}}(k^{\bar{\alpha}},k'^{\bar{\beta}}) =
 \sum_{m=1}^{N^{\bar{\alpha}}}
 \lambda_{(m),I}^{\bar{\alpha} \bar{\beta}} \;
 g_{(m),I}^{\bar{\alpha}}(k^{\bar{\alpha}}) \, g_{(m),I}^{\bar{\beta}}(k'^{\bar{\beta}}).
\end{equation}
Here for convenience new indices of two-body channels $\bar{\alpha}, \bar{\beta} = 1, 2$
were introduced. Correspondence between a two-body index $\bar{\alpha}$ and a pair
$(\alpha, i)$ of three-body channel and Faddeev indices, defining an interacting pair,
can be established with the help of Table I of~\cite{our_KNN_PRC}.
As before, $N^{\bar{\alpha}}$ defines the number of
terms of the potential. As was already stated, we neglected here the $\pi N$ interaction
due to its smallness.

In addition to the coupled-channel potentials (Eq.~(\ref{Vseprb}) with
$\bar{\alpha}, \bar{\beta} > 1$)
we used optical and complex one-channel potentials corresponding to them. Having in
mind, that nowadays many authors misuse the term ``optical'' to a complex
potential, we will call our one-channel potentials ``exact optical''
and ``simple complex''. An exact optical potential by definition reproduces
the elastic part of the coupled-channel interaction exactly. In particular,
the imaginary part of the corresponding amplitude
becomes zero below the lowest channel threshold.

Exact optical one-channel potential, corresponding to a two-channel $V$
(with $N^{\bar{\alpha}} = 1$), is given by equation~(\ref{Vseprb}) with
$\bar{\alpha}, \bar{\beta} = 1$ and the strength parameter defined as
\begin{equation}
\label{lambdaOpt}
\lambda^{11,{\rm Opt}}_{I} = \lambda^{11}_{I} +
\frac{(\lambda^{12}_{I})^2 \,
\langle g^{2}_{I} |\, G_0^{(2)}(z^{(2)}) |\, g^{2}_{I} \rangle}{
1 - \lambda^{22}_{I} \, \langle g^{2}_{I} |\, G_0^{(2)}(z^{(2)}) |\, g^{2}_{I} \rangle
} \,,
\end{equation}
where $\lambda^{\bar{\alpha},\bar{\beta}}_I$ are strength parameters of the two-channel
potential, $| g^{2}_{I} \rangle$ is the form-factor of the second channel.
Having in mind, that two-body free Green's function $G_0^{(2)}$ depends on the corresponding
two-body energy $z^{(2)}$, we see, that $\lambda^{11,{\rm Opt}}_{I}$ of the exact optical
potential is an energy-dependent complex function.
In contrast to it, a strength parameter $\lambda^{11,{\rm Complex}}_{I}$ of a simple
complex potential is a complex constant, therefore, the simple complex potential is
energy independent. The strength parameter of a simple complex potential is chosen in
such a way, that the potential reproduces some characteristics of the full interaction, say,
scattering lengths. Exact optical and simple complex potentials take into account flux
losses into inelastic channels through imaginary parts of the strength parameters.

\subsection{$\bar{K}N - \pi \Sigma$ potential}
\label{KNpiSigma.subsect}

Two versions of phenomenological coupled-channel $\bar{K}N - \pi \Sigma$ potential,
constructed in~\cite{ourPRC_isobreak}, have one- and two-pole form of $\Lambda(1405)$
resonance and simultaneously reproduce all existing experimental data. In the present work
we performed new fits of the experimental data with the same potential forms. The one-term
($N^{\bar{\alpha}}=1$) two-channel
potential is defined by Eq.~(\ref{Vseprb}), where $\bar{\alpha}=1$ denotes $\bar{K}N$,
$\bar{\alpha}=2$ -- $\pi \Sigma$ channel.
All physical values for data fitting were obtained by solving coupled-channel
Lippmann--Schwinger equations with direct inclusion of the Coulomb potential into 
the $K^- p$ system. Another source of isospin symmetry breaking is
the use of physical masses for $K^-, \bar{K}^0, p$ and $n$.

In comparison to~\cite{ourPRC_isobreak}, where one one-pole and one
two-pole potentials were constructed, here we obtained two sets of potential parameters
$\lambda_{I}^{\bar{\alpha} \bar{\beta}}$ and $\beta_{I}^{\bar{\alpha}}$:
one set for one-pole and another set for two-pole structure of $\Lambda(1405)$ resonance.
Each potential of these sets gives medium value for threshold branching
ratios~\cite{gammaKp1,gammaKp2}:
\begin{eqnarray}
\label{gamma}
\gamma &=& 2.36 \pm 0.04  \,, \\
\label{RpiSigma}
R_{\pi \Sigma} &=& 0.709 \pm 0.011 \,,
\end{eqnarray}
where the second is a new ratio, constructed from experimentally measured $R_c$ and $R_n$:
\begin{equation}
 R_{\pi \Sigma} =  \frac{R_c}{1-R_n \, (1 - R_c)} \,.
\end{equation}
In contrast to $R_c$ and $R_n$, the new branching ratio $R_{\pi \Sigma}$  does
not contain the $\pi^0 \Lambda$ channel, which is taken into account in our formalism   
only effectively through the non-zero imaginary part of $\lambda_1^{\bar{K}\bar{K}}$
parameter.

\begin{figure*}
\centering
\includegraphics[width=0.85\textwidth]{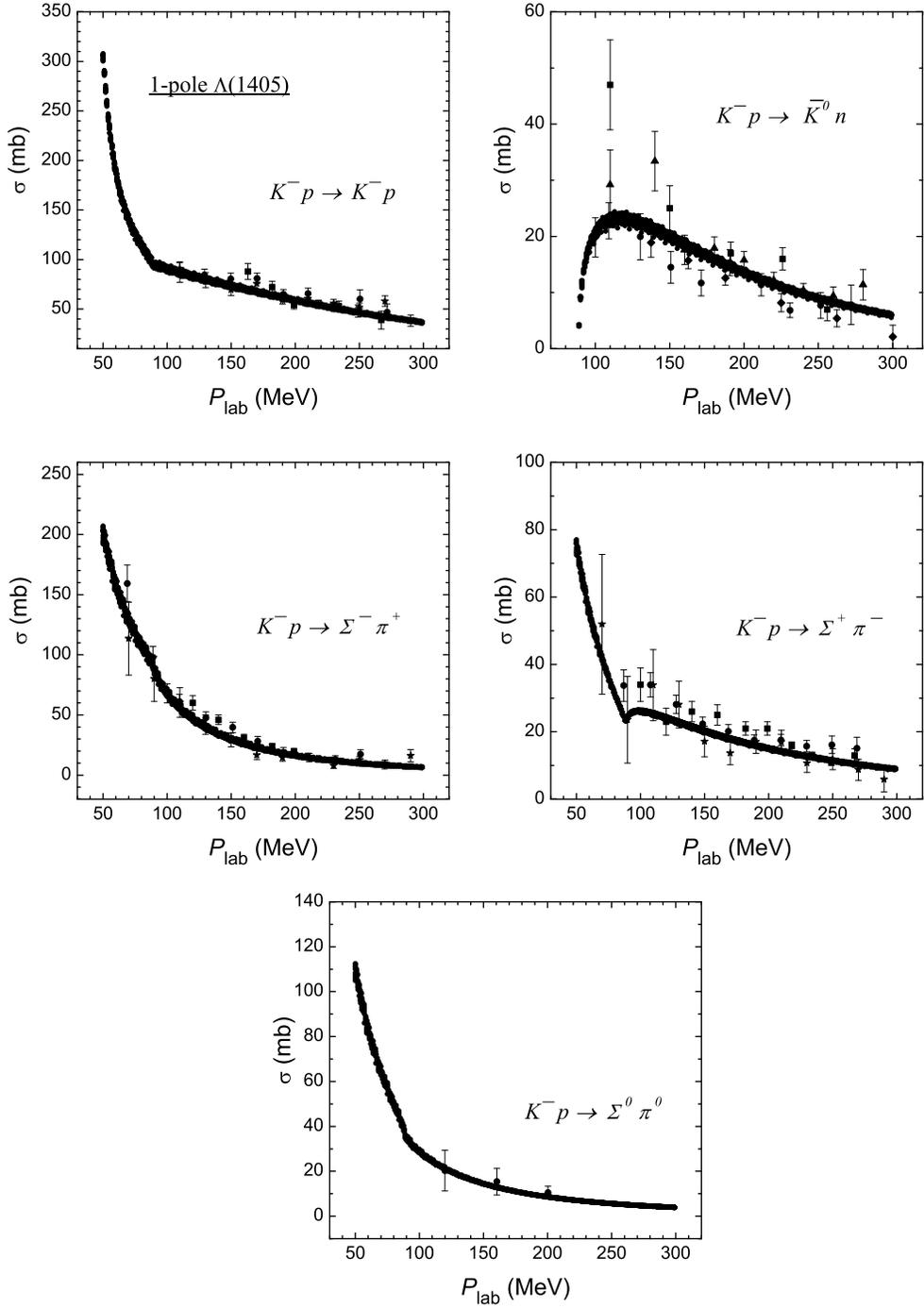}
\caption{Comparison of the elastic and inelastic $K^- p$
cross-sections (filled circles) for the one-pole sets of the
$\bar{K}N - \pi \Sigma$ potential with experimental
data~\protect\cite{Kp2exp,Kp3exp,Kp4exp,Kp5exp,Kp6exp} (data points).
The theoretical bands are formed by all lines obtained with
individual potentials within the set.
\label{Fiveplots_1pole_sets.fig}}
\end{figure*}

\begin{figure*}
\centering
\includegraphics[width=0.85\textwidth]{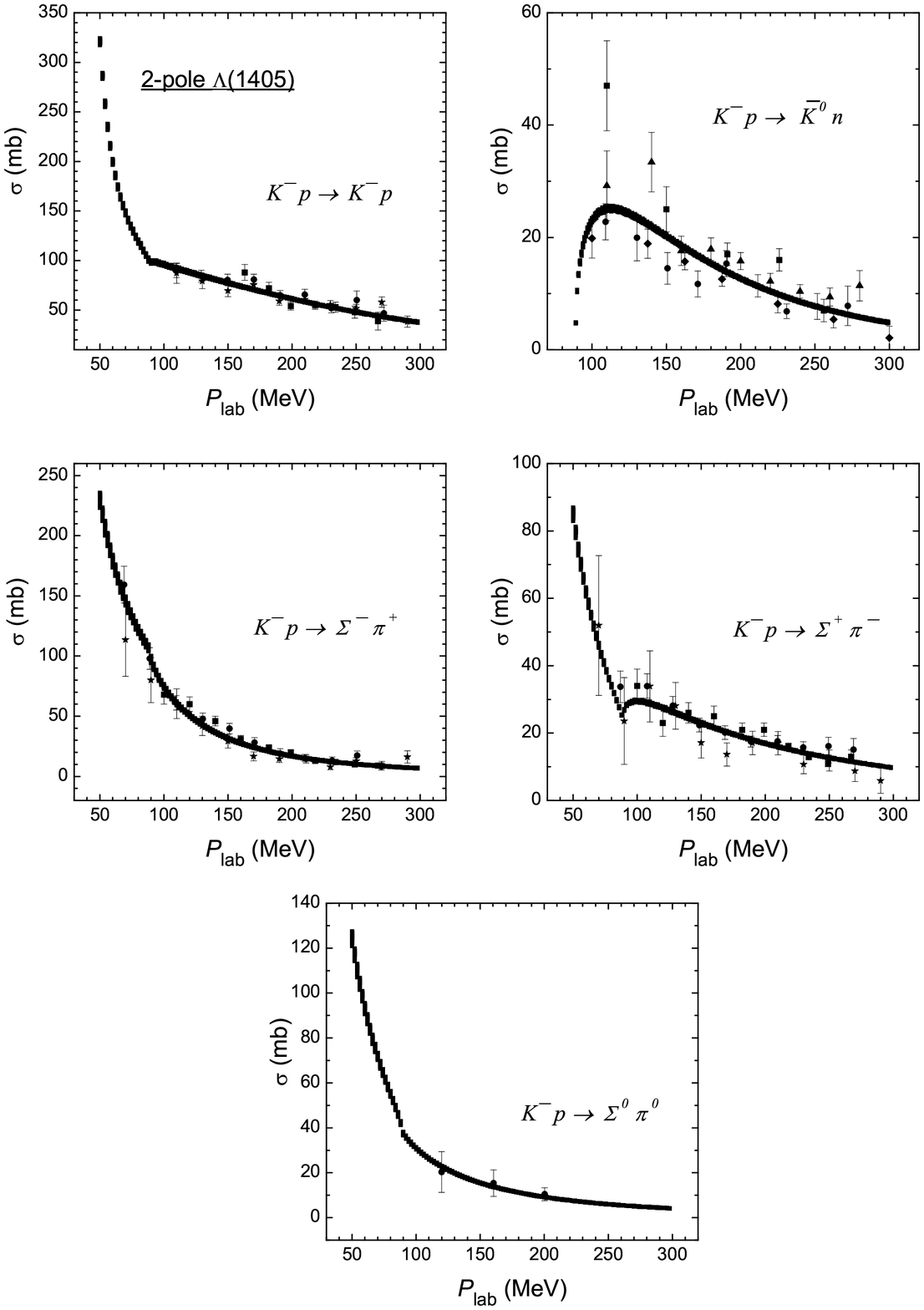}
\caption{Comparison of the elastic and inelastic $K^- p$
cross-sections (filled circles) for the two-pole sets of the
$\bar{K}N - \pi \Sigma$ potential with experimental
data~\protect\cite{Kp2exp,Kp3exp,Kp4exp,Kp5exp,Kp6exp} (data points).
The theoretical bands are formed by all lines obtained with
individual potentials within the set.
\label{Fiveplots_2pole_sets.fig}}
\end{figure*}

Elastic and inelastic $K^- p$ cross-sections $K^- p \to {K}^- p$,
$K^- p \to \bar{K}^0 n$, $K^- p \to \pi^+ \Sigma^-$, $K^- p \to \pi^- \Sigma^+$,
and $K^- p \to \pi^0 \Sigma^0$ are properly reproduced by our new potentials as well.
The theoretical results together with with experimental
data~\cite{Kp2exp,Kp3exp,Kp4exp,Kp5exp,Kp6exp} are shown in
Fig.~\ref{Fiveplots_1pole_sets.fig} and Fig.~\ref{Fiveplots_2pole_sets.fig}  for
one- and two-pole sets of potentials, respectively (we did not take into
consideration data from~\cite{Kp1exp} with huge error bars).
The potentials within each set provide slightly different
cross-sections, which results in a band instead of a line in the figures.
In the same way as in~\cite{ourPRC_isobreak} we defined ``total elastic'' $K^- p$
cross-section as an integral of the differential cross-section over
the $-1 \leq \cos{\theta} \leq 0.966$ region due to the singularity of the pure
Coulomb transition matrix in forward direction.

\begin{figure}
\centering
\includegraphics[width=0.35\textwidth, angle=-90]{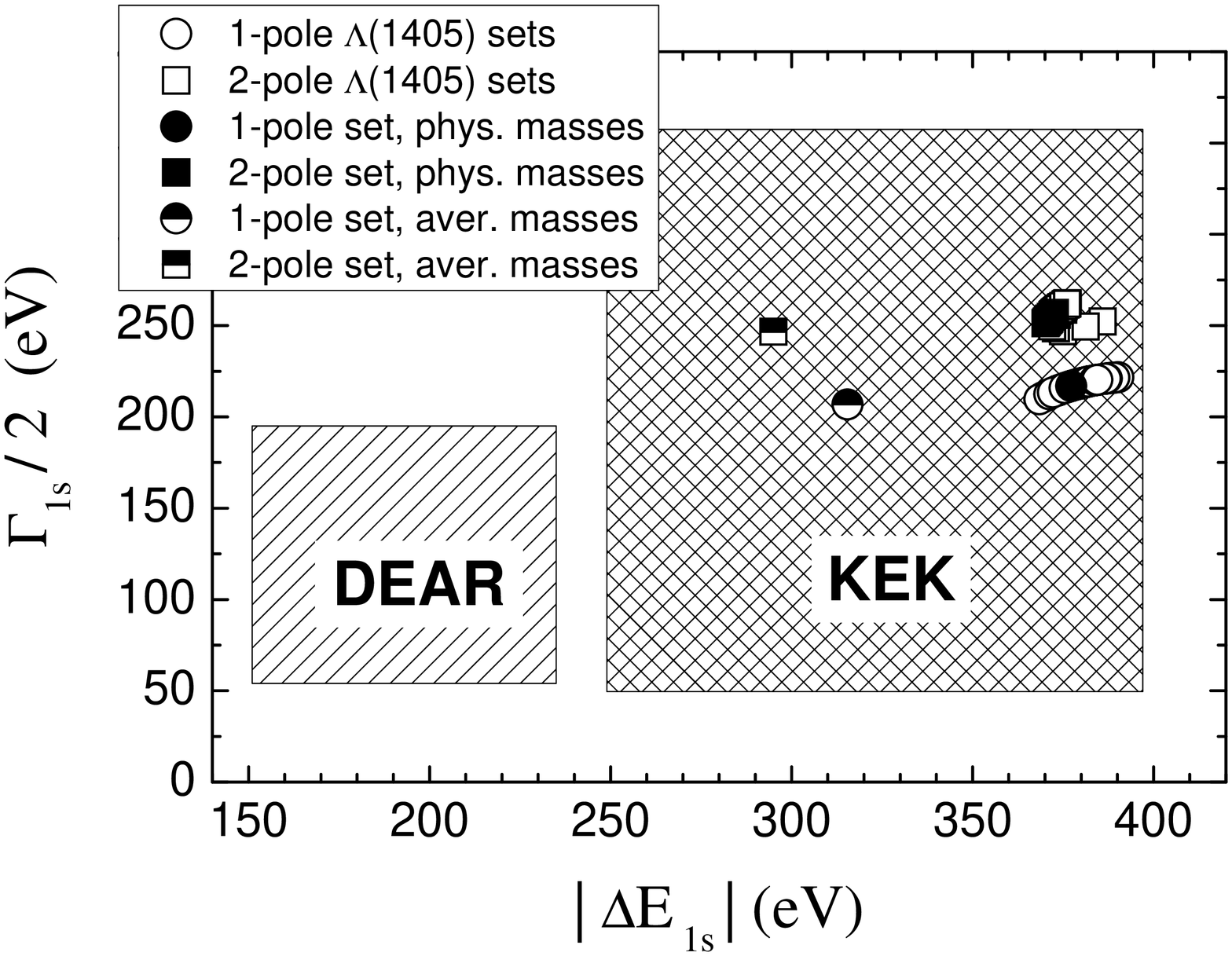}
\caption{Kaonic hydrogen $1s$ level shift $|\Delta E|$ (absolute value)
and width $\Gamma$ values for the one-pole (empty circles) and two-pole
(empty squares) sets of the $\bar{K}N - \pi \Sigma$ potential.
Filled circle and filled square denote the results for the
one- and two-pole representative potentials, correspondingly,
obtained with physical masses. The same kaonic hydrogen observables calculated
with averaged masses are denoted by half-empty circle and square.
Experimental DEAR and KEK $1 \sigma$ confidence regions are also shown.
\label{KEK_DEAR_full.fig}}
\end{figure}

Characteristics of kaonic hydrogen atom $(\Delta E_{1s}, \Gamma_{1s})$ provided by
our potentials are situated within $1 \sigma$ KEK~\cite{KEK1s} experimental region,
see Fig.~\ref{KEK_DEAR_full.fig}. Experimental results obtained
by DEAR collaboration~\cite{DEAR1s} are also shown. 
See \footnote{
After sending the present article
to the journal the results of SIDDHARTA experiment appeared,
see~\cite{SIDDHARTA1s}. The $1 \sigma$ SIDDHARTA region is situated inside KEK square
with comparable width $\Gamma_{1s}$ and smaller level shift $\Delta E_{1s}$ values than 
those provided by our potentials.}
as well.

\begin{figure*}
\centering
\includegraphics[width=0.30\textwidth, angle=-90]{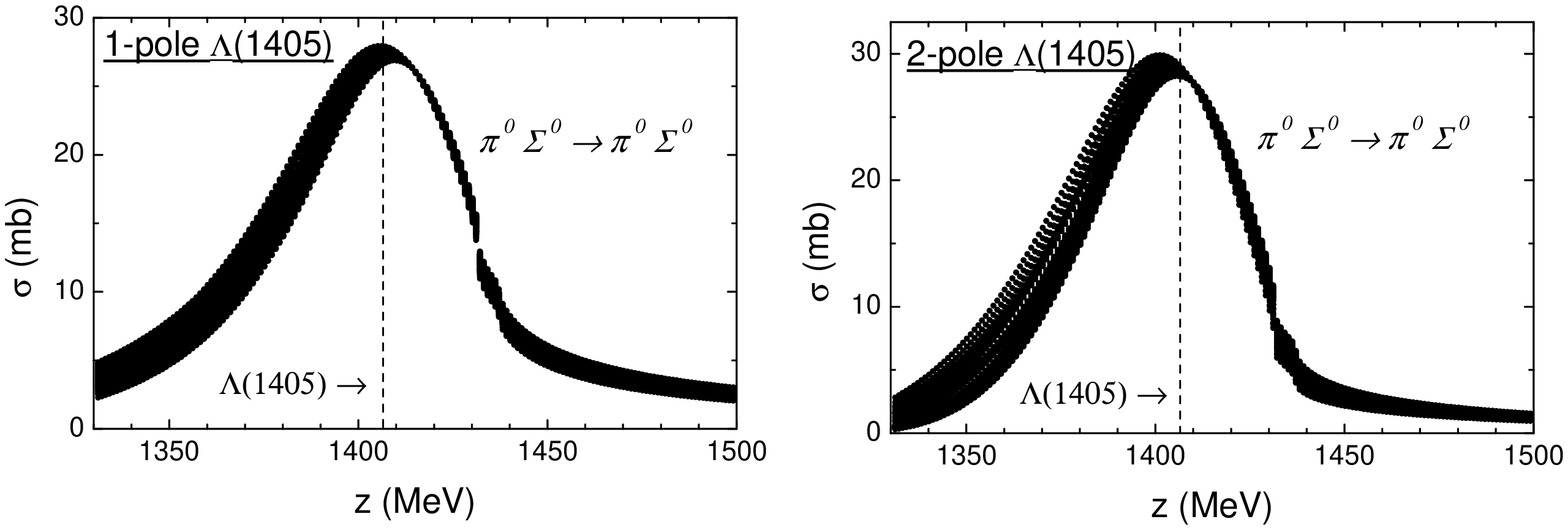}
\caption{Elastic $\pi^0 \Sigma^0$ cross-sections
for the one-pole (left) and two-pole sets of $\bar{K}N - \pi \Sigma$ potential.
The theoretical bands are formed by all lines obtained with
individual potentials within the corresponding set.
\label{pi0Sig0_sets.fig}}
\end{figure*}

Typical resonance behaviour manifests itself in $\pi^0 \Sigma^0$ elastic cross-sections,
corresponding to one- and two-pole sets of $\bar{K}N - \pi \Sigma$ potential,
see Fig.~\ref{pi0Sig0_sets.fig} (bands, consisting of individual lines). All resonance
maxima are situated near Particle data group (PDG)~\cite{PDG} values for the
$\Lambda(1405)$ resonance mass and width
\begin{equation}
M^{PDG}_{\Lambda(1405)} = 1406.5 \pm 4.0 \; {\rm MeV}, \quad
\Gamma^{PDG}_{\Lambda(1405)} = 50 \pm 2.0 \; {\rm MeV}.
\end{equation}
PDG mass of the $\Lambda(1405)$ resonance is also shown at the figures. Strong pole
positions and widths are slightly different for the potentials within one- and two-pole
sets of potentials.
%----------------------------------------------------------------
\begin{center}
\begin{table}[hb]
\caption{Parameters of the representative one- and two-pole $\bar{K}N - \pi \Sigma$ potentials:
range $\beta^{\bar{\alpha}}$ (independent on two-body isospin $I$), strength
$\lambda^{\bar{\alpha} \bar{\beta}}_I$, and additional parameter $s$
of the two-pole model.}
\label{params.tab}
\begin{tabular}{cccccccccc}
\hline \hline \noalign{\smallskip}
{} &\, $\beta^{\bar{K}N}$ &\, $\beta^{\pi \Sigma}$
&\, $\lambda^{\bar{K}\bar{K}}_{0}$ &\, $\lambda^{\bar{K}\pi}_{0}$ &\, $\lambda^{\pi \pi}_{0}$
&\, $\lambda^{\bar{K}\bar{K}}_{1}$ &\, $\lambda^{\bar{K}\pi}_{1}$ &\, $\lambda^{\pi \pi}_{1}$
&\, $s$ \\
\noalign{\smallskip} \hline \noalign{\smallskip}
$V^{\rm one-pole}_{\bar{K}N - \pi \Sigma}$ &\, $3.41$ &\, $1.62$ &\, $-1.2769$ &\, $0.5586$
 &\, $0.2024$ &\, $1.0623 - i \, 0.3251$ &\, $1.8315$ &\, $1.7158$ &\, $0.0000$ \\
$V^{\rm two-pole}_{\bar{K}N - \pi \Sigma}$ &\, $3.72$ &\, $1.00$ &\, $-1.6588$ &\, $0.4672$
 &\, $0.0072$ &\, $0.7329 - i \, 0.2967$ &\, $1.5357$ &\, $1.0744$ &\, $-0.8433$ \\
\noalign{\smallskip} \hline \hline
\end{tabular}
\end{table}
\end{center}
%----------------------------------------------------------------

For a more detailed description of the properties of our models of $\bar{K}N - \pi \Sigma$
interaction we chose two ``representative'' potentials: one with one-pole and another
with two-pole structure of $\Lambda(1405)$ resonance. Parameters of the potentials
are shown in Table~\ref{params.tab}, the corresponding observables  - in
Table~\ref{phys_char1.tab}. The latter contains the above mentioned $\gamma$, $R_{\pi \Sigma}$
ratios and kaonic hydrogen characteristics $\Delta E_{1s}$, $\Gamma_{1s}$ together with
positions of the strong poles $z_1$ and $z_2$. Potentials with equal $z_1$ values were
chosen as representative ones. The strong $K^- p$ scattering lengths $a_{K^- p}$, exactly
corresponding to the kaonic hydrogen observables, are shown in Table~\ref{phys_char1.tab}.
Since both $a_{K^- p}$ and $(\Delta E_{1s},\Gamma_{1s})$ were obtained by exact
solution of Lippmann-Schwinger equation, the relation between them does not correspond
to any commonly used approximate formula.
From Figs.~\ref{Fiveplots_1pole_sets.fig}, \ref{Fiveplots_2pole_sets.fig},
\ref{KEK_DEAR_full.fig} and Table~\ref{phys_char1.tab} it is seen, that our
new one- and two-pole $\bar{K}N - \pi \Sigma$ potentials reproduce all experimental
data within experimental errors indistinguishably well in the same way as the 
ones in~\cite{ourPRC_isobreak}. Therefore it is not possible to give preference
to one of the two versions.
%----------------------------------------------------------------
\begin{center}
\begin{table}[ht]
\caption{Physical characteristics of the representative one-pole and two-pole potentials
(full version with physical masses): strong pole(s) position(s) $z_1$ (and $z_2$), kaonic hydrogen
$1s$ level shift $\Delta E_{1s}$ and width $\Gamma_{1s}$, $K^- p$ scattering length $a_{K^- p}$,
and threshold branching ratios $\gamma$ and $R_{\pi \Sigma}$.}
\label{phys_char1.tab}
\begin{tabular}{ccc}
\hline \hline \noalign{\smallskip}
 & $V^{\rm one-pole}_{\bar{K}N - \pi \Sigma}$ & $V^{\rm two-pole}_{\bar{K}N - \pi \Sigma}$ \\
\noalign{\smallskip} \hline \noalign{\smallskip}
$z_1$ (MeV) & \qquad $1409 - i 36$  \qquad & \qquad
       $1409 - i 36$ \qquad\\
$z_2$ (MeV) & \qquad $-$  \qquad & \qquad
       $1381 - i 105$ \qquad\\
$\Delta E_{1s}$ (eV)   & $-377$ &  $-373$ \\
$\Gamma_{1s}$ (eV)     & $434$ &  $514$  \\
$a_{K^- p}$ (fm)  & $-1.00 + i 0.68$ & $-0.96 + i 0.80$ \\
$\gamma$   & $2.36$ & $2.36$ \\
$R_{\pi \Sigma}$  & $0.709$ & $0.709$ \\
\noalign{\smallskip} \hline \hline
\end{tabular}
\end{table}
\end{center}
%----------------------------------------------------------------

\begin{figure*}
\centering
\includegraphics[width=0.30\textwidth, angle=-90]{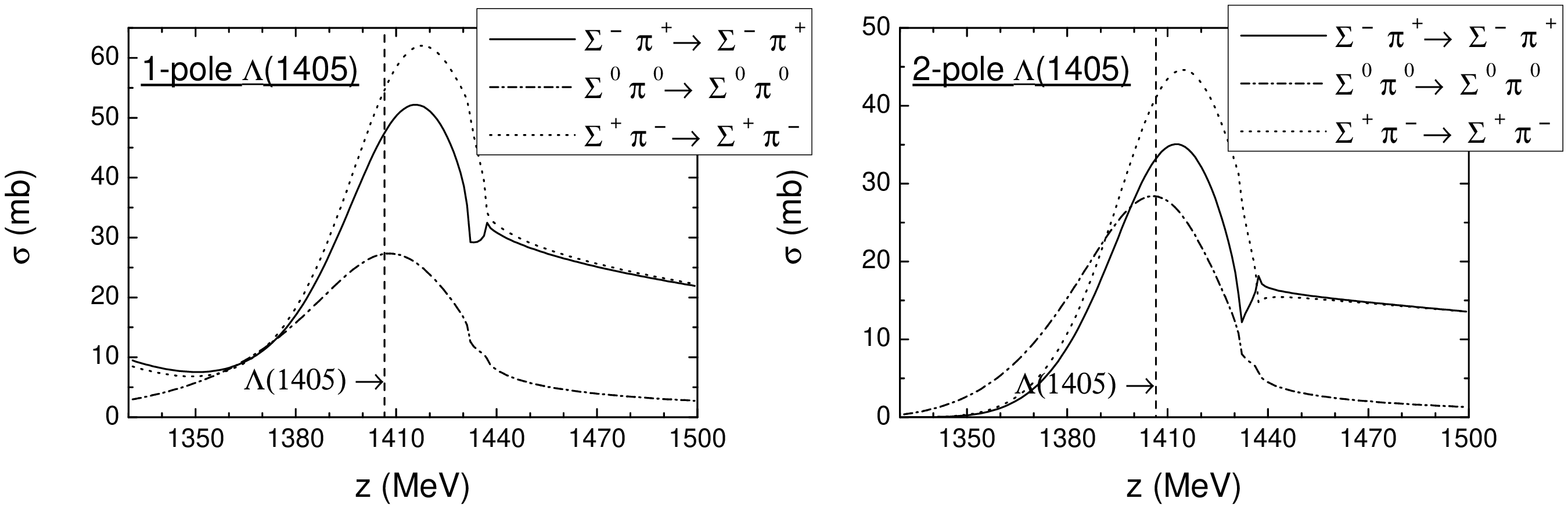}
\caption{Three charged elastic $\pi \Sigma$ cross-sections for the representative
one-pole (left) and two-pole $\bar{K}N - \pi \Sigma$ potentials.
\label{Lambda1405_charged.fig}}
\end{figure*}

In addition, we checked several arguments, which were presented in support 
to the idea of the two-pole structure of $\Lambda(1405)$ resonance.
One of them is a difference between charged $\pi \Sigma$ cross-sections,
which is seen in different experiments,
such as CLAS~\cite{CLAS}. In order to check the assumption, that the difference
is caused by the two-pole structure, we plotted $\pi^+ \Sigma^-$, $\pi^- \Sigma^+$,
and $\pi^0 \Sigma^0$ elastic cross-sections. The result is seen in
Fig.~\ref{Lambda1405_charged.fig}: the cross-sections are different and
their maxima are shifted one from another for both one- and two-pole versions
of $\bar{K}N - \pi \Sigma$ potential. Therefore, it is not a proof
of the two-pole structure, but a manifestation of an isospin-breaking effect
and different background.

\begin{figure}
\centering
\includegraphics[width=0.85\textwidth]{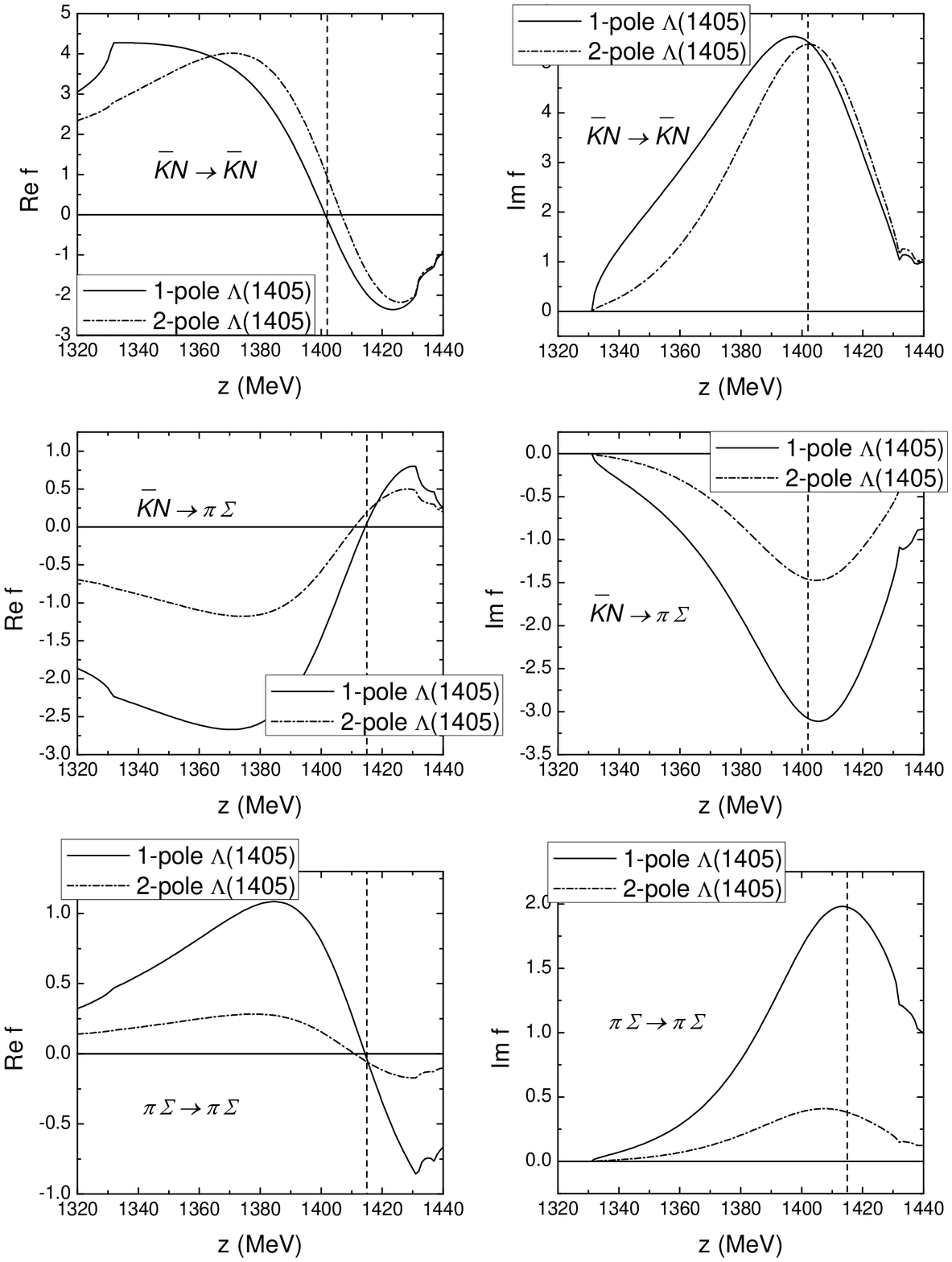}
\caption{Real (left column) and imaginary (right column) parts of the
$\bar{K}N - \bar{K}N$, $\bar{K}N - \pi \Sigma$ and $\pi \Sigma - \pi \Sigma$
amplitudes obtained with one-pole (solid line) and two-pole (dash-dotted line)
representative $\bar{K}N - \pi \Sigma$ potentials.
The energies at which real parts of the one-pole amplitudes cross real axes
are denoted by vertical dashed lines.
\label{amplitudes.fig}}
\end{figure}

Another argument for two-pole structure comes from the fact, that the poles in
a two-pole model are coupled to different channels. Indeed, gradually switching
off the coupling between $\bar{K}N$ and $\pi \Sigma$ channels turns  
the highest of the poles into a real bound state in $\bar{K}N$, while  
the lowest one become a resonance in uncoupled $\pi \Sigma$ channel.
Such a behaviour was demonstrated in several papers, see e.g.~\cite{ourPRC_isobreak}.
Accordingly, it was suggested, that the poles of a two-body model
manifest  themselves in different reactions, in particular, $\bar{K}N - \bar{K}N$,
$\bar{K}N - \pi \Sigma$, and $\pi \Sigma - \pi \Sigma$ amplitudes should
``feel'' only one of the two poles. We checked the hypothesis, the results
are demonstrated in Fig.~\ref{amplitudes.fig}. Indeed,
real parts of $\bar{K}N - \bar{K}N$, $\bar{K}N - \pi \Sigma$, and
$\pi \Sigma - \pi \Sigma$ amplitudes in $I=0$ state
cross real axis at different energies, but it is true for both versions
of the potential. In fact, the difference is even larger for the one-pole amplitudes.
In our opinion, the effect is caused by different background contributions in the
reactions independently of the number of poles.
Consequently, a proof of the two-pole structure of the $\bar{K}N - \pi \Sigma$
interaction does not exist.

\begin{figure*}
\centering
\includegraphics[width=0.85\textwidth]{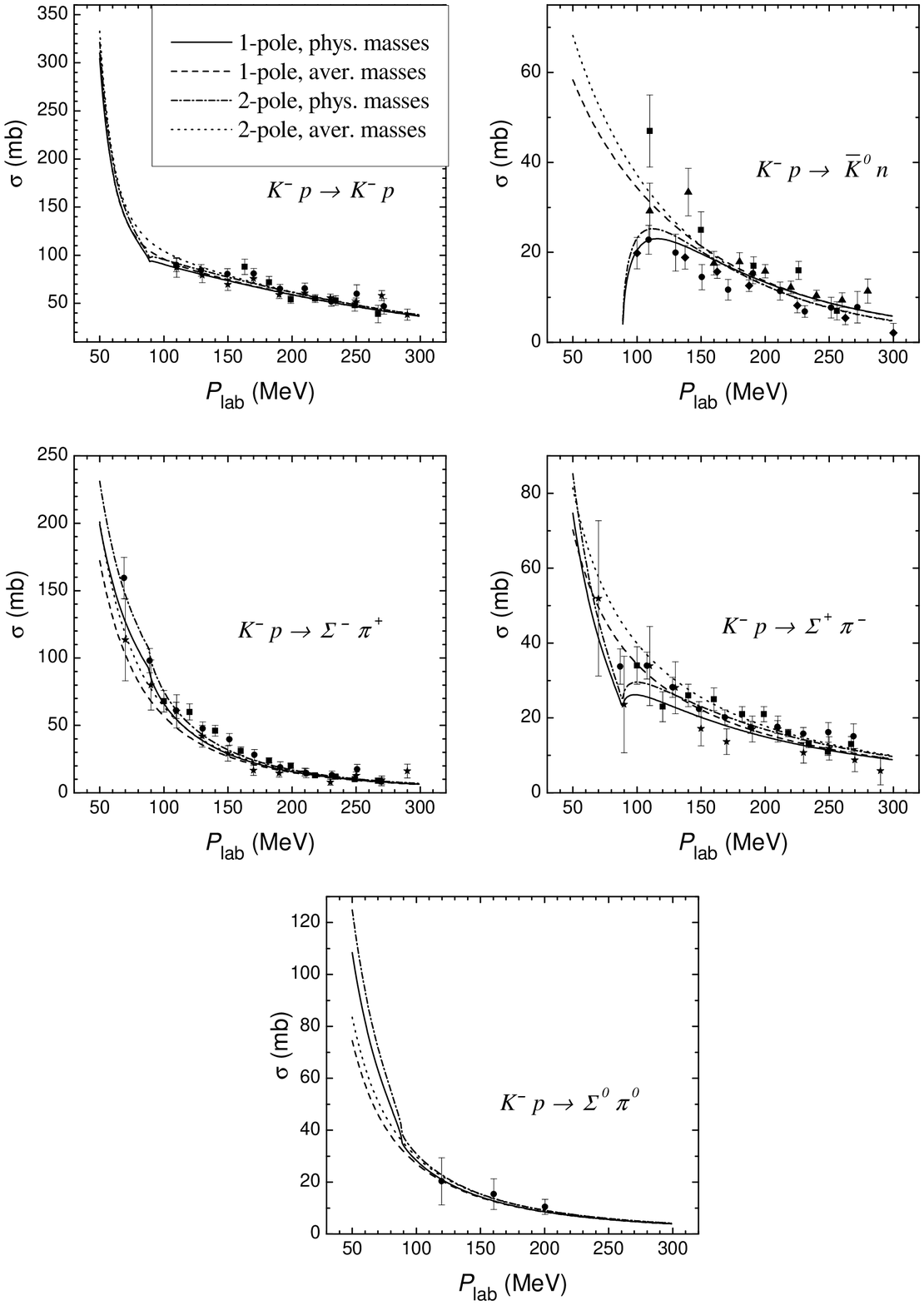}
\caption{Comparison of the theoretical $K^- p$ cross-sections
for the representative one- and two-pole $\bar{K}N - \pi \Sigma$
potentials with experimental
data~\protect\cite{Kp2exp,Kp3exp,Kp4exp,Kp5exp,Kp6exp} (data points).
The results obtained with the physical masses (solid, dash-dotted line) and
the averaged masses (dashed and dotted line) are presented.
\label{Fiveplots_phys_aver.fig}}
\end{figure*}

Coulomb interaction, directly included into two-body Lippmann-Schwinger equations,
was necessary for reproducing experimental data on kaonic hydrogen atom. However,
in the $K^- d$ scattering length calculations it is expected to play
a minor role and can be omitted. We also neglected the difference between physical
masses in isodoublets for $K^- d$ system. The physical characteristics of
$\bar{K}N - \pi \Sigma$ system,
calculated with isospin-averaged masses for $\bar{K}$ and $N$ using the obtained sets
of $\lambda_{I}^{\bar{\alpha} \bar{\beta}}$, $\beta_{I}^{\bar{\alpha}}$ parameters
are shown in Fig.~\ref{KEK_DEAR_full.fig}, Fig.~\ref{Fiveplots_phys_aver.fig}, and
Table~\ref{phys_char2.tab}.
%----------------------------------------------------------------
\begin{center}
\begin{table}[ht]
\caption{The same as in Table~\ref{phys_char1.tab}, but with averaged masses
of the particles.}
\label{phys_char2.tab}
\begin{tabular}{ccc}
\hline \hline \noalign{\smallskip}
 & $V^{\rm one-pole}_{\bar{K}N - \pi \Sigma}$ & $V^{\rm two-pole}_{\bar{K}N - \pi \Sigma}$ \\
\noalign{\smallskip} \hline \noalign{\smallskip}
$z_1$ (MeV) & \qquad $1409 - i 36$  \qquad & \qquad
       $1409 - i 36$ \qquad\\
$z_2$ (MeV) & \qquad $-$  \qquad & \qquad
       $1381 - i 105$ \qquad\\
$\Delta E_{1s}$ (eV)   & $-316$ &  $-295$ \\
$\Gamma_{1s}$ (eV)     & $414$ &  $491$  \\
$a_{K^- p}$ (fm)  & $-0.80 + i 0.62$ & $-0.72 + i 0.73$ \\
$\gamma$   & $4.18$ & $4.54$ \\
$R_{\pi \Sigma}$  & $0.761$ & $0.768$ \\
\noalign{\smallskip} \hline \hline
\end{tabular}
\end{table}
\end{center}
%----------------------------------------------------------------

``Averaged'' points $(\Delta E_{1s},\Gamma_{1s})$ for the one- and
two-pole representative potentials in Fig.~\ref{KEK_DEAR_full.fig} 
are shifted to the smaller $|\Delta E_{1s}|$ values relative to the
``physical'' ones. However, they remain inside
$1 \sigma$ KEK region. Fig.~\ref{Fiveplots_phys_aver.fig} demonstrates
``averaged'' and ``physical'' cross-sections, where ``averaged'' ones
naturally do not show threshold behaviour at laboratory momentum $P_{\rm \,lab}$,
corresponding to $\bar{K}^0 n$ threshold. However, differences between
``averaged'' and ``physical'' cross-sections are visible only in the near-threshold
region, where there is no reliable experimental data.

Finally, we see by comparing Table~\ref{phys_char2.tab} with Table~\ref{phys_char1.tab},
that strong pole positions remain almost unchanged. Scattering length
$a_{K^- p}$ changes for both versions of the potential mainly due to the confluence
of the $K^- p$ and $\bar{K}^0 n$ thresholds into one $\bar{K}N$ threshold.
Accordingly, threshold
branching ratios $\gamma$~(\ref{gamma}) and $R_{\pi \Sigma}$~(\ref{RpiSigma})
are the only observables, which are considerably changed after introducing
isospin-averaged masses instead of physical ones.

\subsection{Nucleon-nucleon potentials}
\label{NN.subsect}

Antisymmetrized three-body equations for $K^- d$ system with $s$-wave interactions
contain only spin-triplet $NN$ interaction.
We used different $NN$ potentials in order to investigate
dependence of the $K^- d$ scattering length on nucleon-nucleon interaction models.
One of them is a two-term separable $NN$ potential~\cite{DolesNN},
which reproduces Argonne $V18$~\cite{ArgonneV18} phase shifts and, therefore, is
repulsive at short distances. The potential, which will be called 
TSA, is described by Eq.(\ref{Vseprb}) with $N^{\bar{\alpha}} = 2$ and
$\bar{\alpha} = \bar{\beta} = 1$ (the $NN$ interaction is obviously is diagonal
in particle indices). Two versions of the potential
(TSA-A and TSA-B) with slightly different form-factors were used:
\begin{eqnarray}
&{}& g_{(m)}^{A,NN}(k) = \sum_{n=1}^2 \frac{\gamma_{(m)n}^A}{(\beta_{(m)n}^A)^2 + k^2},
\quad {\rm for \;} (m)=1,2 \\
\nonumber
&{}& g_{(1)}^{B,NN}(k) = \sum_{n=1}^3 \frac{\gamma_{(1)n}^B}{(\beta_{(1)n}^B)^2 + k^2},
\quad
g_{(2)}^{B,NN}(k) = \sum_{n=1}^2 \frac{\gamma_{(2)n}^B}{(\beta_{(2)n}^B)^2 + k^2}.
\end{eqnarray}
TSA-A and TSA-B potentials in the ${}^3S_1$ state yield the following scattering lengths
and effective radii
\begin{eqnarray}
a^A(np) = -5.402 {\, \rm fm}, \, r_{\rm eff}^A(np) = 1.754 {\, \rm fm}, \\
a^B(np) = -5.413 {\, \rm fm}, \, r_{\rm eff}^B(np) = 1.760 {\, \rm fm},
\end{eqnarray}
and give correct binding energy of the deuteron $E_{\rm{deu}} = -2.2246$ MeV.

We also used one-term PEST potential (Eq.~(\ref{Vseprb}) with
$N^{\bar{\alpha}} = 1$) from Ref.~\cite{NNpot}, which is a separabelization of
the Paris model of $NN$ interaction. The
strength parameter of PEST $\lambda=-1$, the form-factor is defined by
\begin{equation}
g_{I}^{NN}(k) = \frac{1}{2 \sqrt{\pi}} \, \sum_{n=1}^6
\frac{c_{n,I}^{NN}}{k^2 + (\beta_{n,I}^{NN})^2} \, ,
\end{equation}
where the constants $c_{n,I}^{NN}$ and $\beta_{n,I}^{NN}$ are listed
in Ref.~\cite{NNpot}. PEST is equivalent to the Paris
potential on and off energy shell up to $E_{\,\rm lab} \sim 50$ MeV. 
It reproduces the deuteron binding
energy $E_{\,\rm deu} = -2.2249$ MeV, as well as the triplet and singlet $NN$
scattering lengths, $a(\,{}^3S_1) = -5.422$ fm and $a(\,{}^1S_0) = 17.534$ fm,
respectively.

\begin{figure}
\centering
\includegraphics[width=0.35\textwidth, angle=-90]{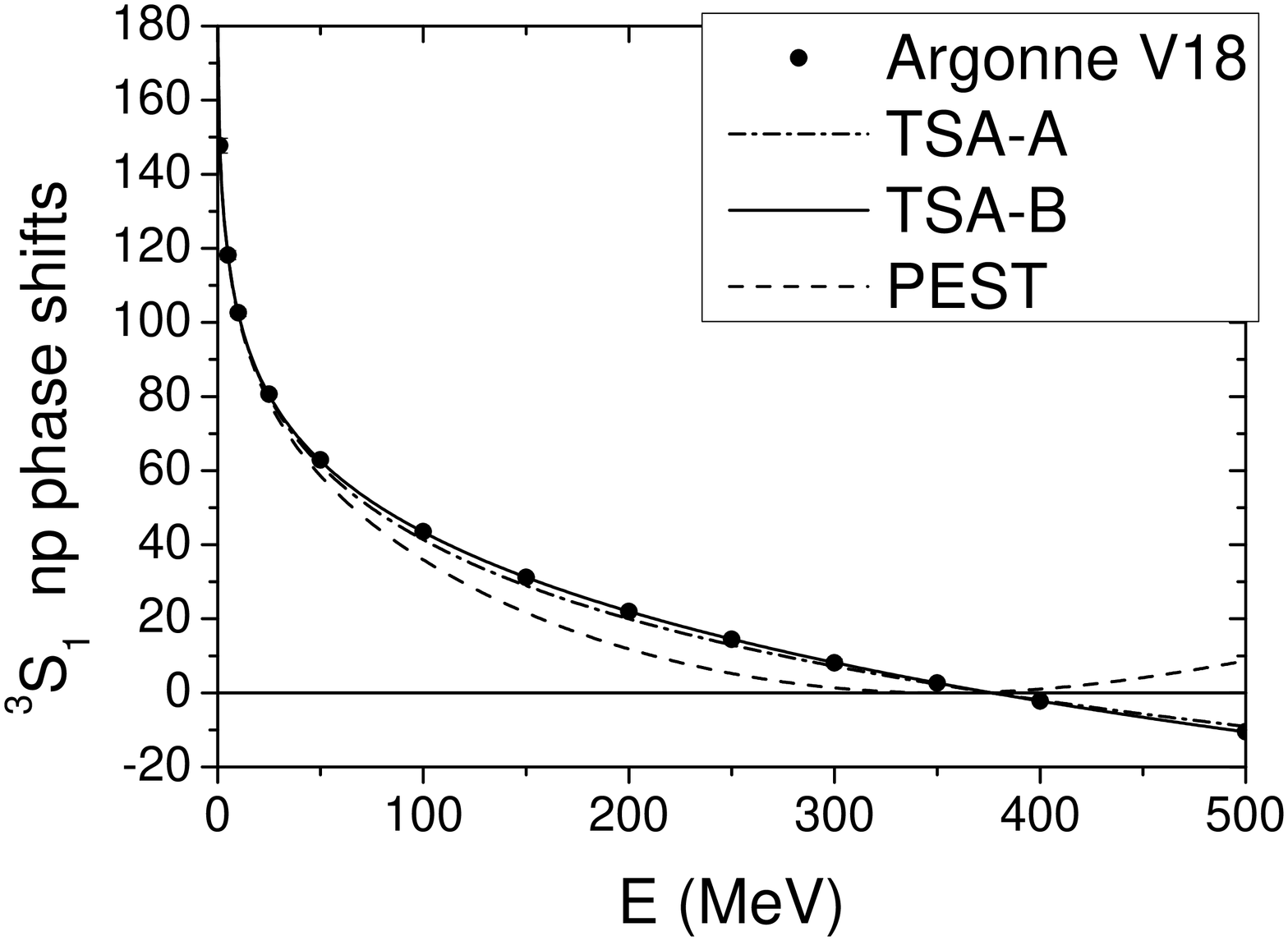}
\caption{${}^3S_1$ np phase shifts of TSA-A (dash-dotted line),
TSA-B (solid line), and PEST (dashed line) $NN$ potentials
in comparison with those of Argonne $v18$ potential (solid circles).
\label{Doles_NN}}
\end{figure}

The ${}^3S_1$ phase shifts for the three $NN$ potentials
are shown in Fig.~\ref{Doles_NN} together with characteristics of the Argonne V18
model. Almost indistinguishable lines correspond to the two-term TSA-A and TSA-B
potentials, which are very good at reproducing the Argonne V18 phase shifts.
Their crossing of the real axis is a consequence of repulsion at short distances.
One-term PEST $NN$ potential does not have such a property, but at low energies
its phase shifts are also close to the ``etalon'' ones.
Therefore, the two-term $NN$ potentials, increasing the
number of equations in a three-body system, reproduce
properties of $NN$ interaction better than the one-term potential.

\subsection{$\Sigma N -\Lambda N$ interaction}
\label{SigmaN.subsect}

The $\Sigma N$ interaction, which is coupled with $\Lambda N$ channel
in $I = \frac{1}{2}$ isospin state,
is usually assumed to be spin-dependent~\cite{SigmaNth1,SigmaNth2}.
We constructed new versions of $\Sigma N - \Lambda N$ potential in such a way,
that it reproduces existing experimental
data~\cite{SigmaN1,SigmaN2,SigmaN3,SigmaN4,SigmaN5}. One-term separable
potentials, described by Eq.(\ref{Vseprb}) with $N^{\bar{\alpha}}=1$
and Yamaguchi form-factors
\begin{equation}
g_{I,S}^{\Sigma N}(k) = \frac{1}{k^2 + (\beta_{I,S}^{\Sigma N})^2}
\end{equation}
were used for the two possible isospin states. But the number of channels
is different for $I=\frac{1}{2}$ and $I=\frac{3}{2}$.
The parameters of the one-channel ($\bar{\alpha} = \bar{\beta} = 1$)
$\Sigma N$ interaction with isospin $I=\frac{3}{2}$
were fitted to the $\Sigma^+ p \to \Sigma^+ p$ cross-sections. Isospin
one-half $\Sigma N$ is coupled to $\Lambda N$ channel, therefore,
at first we constructed a coupled-channel potential of the $I=\frac{1}{2}$
$\Sigma N - \Lambda N$ interaction. The channel indices
$\bar{\alpha}, \bar{\beta} = 1,2$ in (\ref{Vseprb}) denote
$\Sigma N$ and $\Lambda N$ channel, correspondingly.
The coupled-channel $I=\frac{1}{2}$ potential together with the
one-channel $I=\frac{3}{2}$ potential reproduces the $\Sigma^- p \to \Sigma^- p$,
$\Sigma^- p \to \Sigma^0 n$, $\Sigma^- p \to \Lambda n$, and
$\Lambda p \to \Lambda p$ cross-sections.
%----------------------------------------------------------------
\begin{center}
\begin{table}[ht]
\caption{Range $\beta^{\bar{\alpha}}$ (independent on two-body
isospin $I$) and strength $\lambda^{\bar{\alpha} \bar{\beta}}_I$
parameters of the two $\Sigma N - \Lambda N$ potentials: $V_{S}^{\rm Sdep}$
and $V^{\rm Sind}$ ($S$ stands for the spin). Scattering lengths
$a^{\bar{\alpha}}_{I}$ of $\Sigma N$ and $\Lambda N$ systems are also shown (in fm).}
\label{paramsSNLN.tab}
\begin{tabular}{cccccccccc}
\hline \hline \noalign{\smallskip}
{} &\, $\beta^{\Sigma N}$ &\, $\lambda^{\Sigma N}_{\frac{3}{2}}$
&\, $\beta^{\Lambda N}$ &\, $\lambda^{\Sigma \Sigma}_{\frac{1}{2}}$
&\, $\lambda^{\Sigma \Lambda}_{\frac{1}{2}}$ &\, $\lambda^{\Lambda \Lambda}_{\frac{1}{2}}$
&\, $a^{\Sigma N}_{\frac{1}{2}}$ &\, $a^{\Sigma N}_{\frac{3}{2}}$
&\, $a^{\Lambda N}_{\frac{1}{2}}$ \\
\noalign{\smallskip} \hline \noalign{\smallskip}
$V_{S=0}^{\rm Sdep}$ &\, $1.25$ &\, $-0.0244$ &\, $0.62$ &\, $-1.9956$
 &\, $1.1408$ &\, $-0.7148$ &\, $-1.90 + i \, 0.08 $ &\, $3.18$
 &\, $1.26$ \\
$V_{S=1}^{\rm Sdep}$ &\, $0.50$ &\, $-0.0007$ &\, $1.03$ &\, $-0.0008$
 &\, $0.0185$ &\, $0.0000$ &\, $-3.17 + i \, 1.30$
 &\, $1.63$ &\, $1.57$ \\
$V^{\rm Sind}$ &\, $0.74$ &\, $-0.0032$ &\, $0.74$ &\, $-0.0011$
 &\, $0.0254$ &\, $0.0190$ &\, $-2.40 + i \, 0.85$
 &\, $1.95$ &\, $-1.47$ \\
\noalign{\smallskip} \hline \hline
\end{tabular}
\end{table}
\end{center}
%----------------------------------------------------------------

\begin{figure*}
\centering
\includegraphics[width=0.85\textwidth]{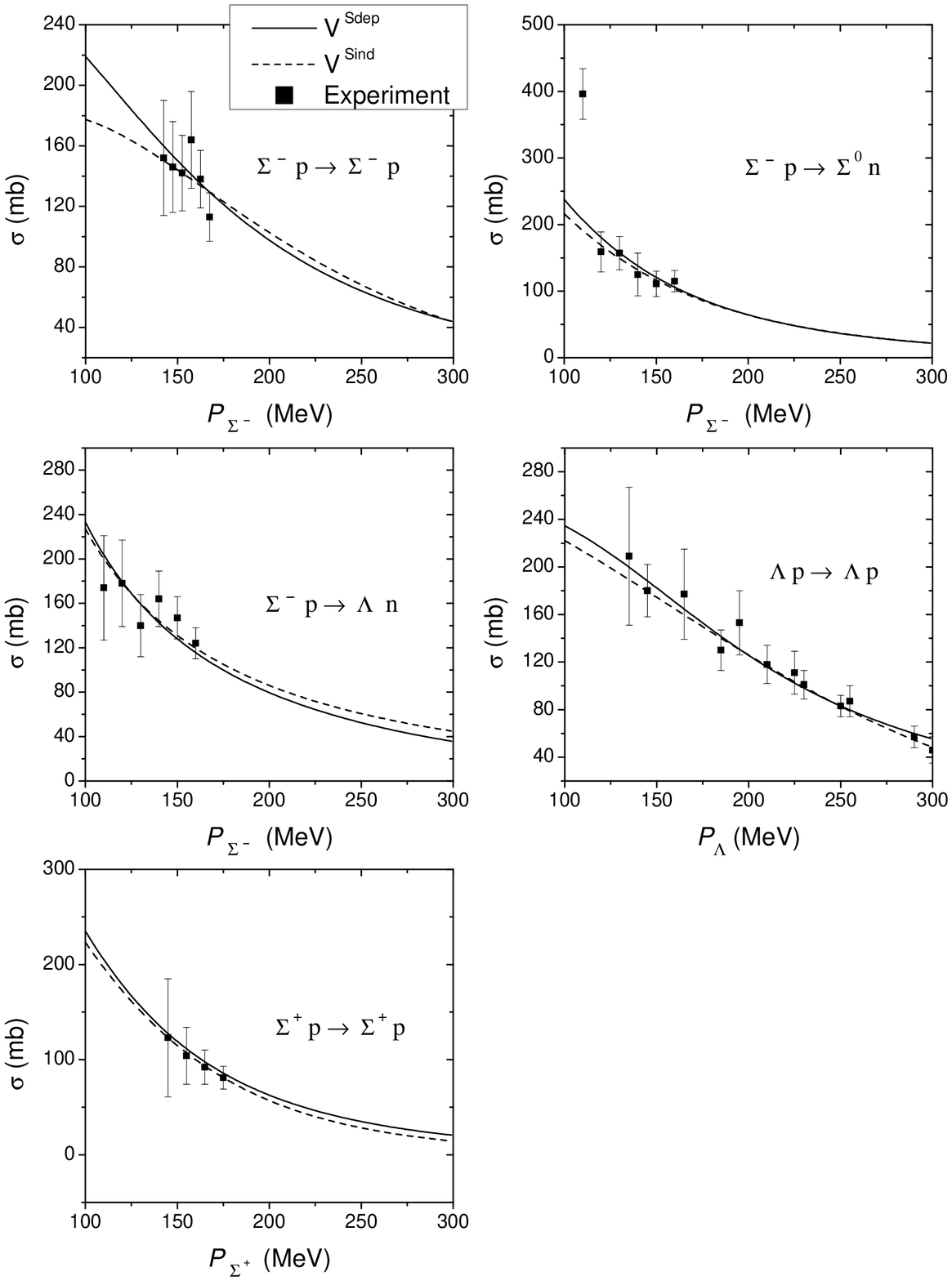}
\caption{Comparison of several theoretical cross-sections
for the spin-dependent (solid line) and spin-independent (dashed line)
coupled-channel $\Sigma N - \Lambda N$ potentials with experimental
data~\protect\cite{SigmaN1,SigmaN2,SigmaN3,SigmaN4,SigmaN5} (data points).
\label{T_SigNLamN.fig}}
\end{figure*}

Two versions of $I=\frac{1}{2}$ $\Sigma N - \Lambda N$ and $I=\frac{3}{2}$ $\Sigma N$
potentials were constructed: one is spin dependent $V^{\rm Sdep}$, the other
$V^{\rm Sind}$ is independent of spin.
Both perfectly reproduce all existing experimental
data~\cite{SigmaN1,SigmaN2,SigmaN3,SigmaN4,SigmaN5} on $\Sigma N$ and
$\Lambda N$ cross-sections, as is seen in Fig.~\ref{T_SigNLamN.fig}. 
Parameters of the potentials
are shown in Table~\ref{paramsSNLN.tab} together with scattering length
values $a^{\Sigma N}_{\frac{1}{2}}$, $a^{\Sigma N}_{\frac{3}{2}}$, and
$a^{\Lambda N}_{\frac{1}{2}}$. The scattering lengths of spin dependent
potential $V^{\rm Sdep}$ are in qualitative agreement
with those provided by more complicated models of $\Sigma N$
interaction~\cite{SigmaNth1,SigmaNth2}. The only exception is $a^{\Sigma N}_{3/2}$
with $S=1$, having opposite sign, which, however, is the same as that given
in previous versions
of the same advanced potentials (our definition of the sign of a scattering length
is opposite to those, used in the mentioned articles).
The scattering lengths, of the spin independent potential $V^{\rm Sind}$, 
are not in such a good agreement, but having in mind, that the scattering length
is not a directly measurable quantity,
we do not consider this difference as a serious defect.

For the three-body $K^- d$ calculations, however, where a channel containing $\Lambda$
is not included directly, we need not a coupled-channel, but a one-channel $\Sigma N$
model of interaction in $I=\frac{1}{2}$ state. Due to this, we additionally constructed
an exact
optical $V^{\Sigma N, \rm{Opt}}$ and simple complex $V^{\Sigma N, \rm{Complex}}$ potentials,
corresponding to the obtained $I=\frac{1}{2}$ $\Sigma N - \Lambda N$ potential.
As was discussed at the beginning of the Section~\ref{two-bodyInp.sect},
the exact optical potential has an energy dependent strength parameter defined by
Eq.~(\ref{lambdaOpt}) and exactly reproduces the elastic $\Sigma N$ amplitude
of the corresponding two-channel potential.
Parameters of the $V^{\Sigma N, \rm{Complex}}$
were found in such a way, that the simple complex potential 
gives the same scattering lengths, as the two-channel potential.
Thus, the exact optical and the simple
complex $\Sigma N (- \Lambda N)$ potentials in $I=\frac{1}{2}$ state and the
one-channel $\Sigma N$ potential in $I=\frac{3}{2}$ were used  during three-body calculations.
The second channel in brackets $(- \Lambda N)$ underlines, that the one-channel potentials
correspond to the coupled-channel one.

\section{Approximate methods}
\label{approximate.sect}

Apart from the full coupled-channel calculation, we performed checks of several
approximate methods, usually used for $K^- d$ scattering length problem, as well.
It is obvious, that a comparison between the full and approximate results is
meaningful only if it is performed with equal two-body input.

\subsection{One-channel AGS calculations}
\label{one-channel.subsect}

In order to investigate the importance of direct inclusion of $\pi \Sigma N$
channel we performed one-channel AGS calculations as well. It means, that
we solved Eq.~(\ref{full_oper_eq}) with $\alpha = \beta = 1$, thus,
only $\bar{K}N$ and $NN$ $T$-matrices enter the equations. We constructed
the exact optical and two simple
complex one-channel $\bar{K}N (- \pi \Sigma)$ potentials approximating the full
coupled-channel one- and two-pole models of interaction. As mentioned at the beginning
of Section~\ref{two-bodyInp.sect}, the exact optical potential $V^{\bar{K}N, \rm{Opt}}$
provides exactly the same elastic $\bar{K}N$ amplitude as the coupled-channel
model of interaction. Its energy-dependent strength parameters are defined by
Eq.~(\ref{lambdaOpt}) with $\bar{\alpha},\bar{\beta} = 1,2$ stands for $\bar{K}N$
and $\pi \Sigma$ channels, correspondingly.

For the simple complex potentials we used range parameters $\beta^{\bar{K}N}$
of the coupled-channel $\bar{K}N - \pi \Sigma$ models of interaction.
The complex $\lambda^{11, \rm{Complex}}_{I}$ constants were obtained in two ways.
The first version of the simple complex $\bar{K}N$ potential
$V^{\bar{K}N, \rm{Complex}}_{(a,z)}$ reproduces $K^- p$
scattering length $a_{K^- p}$ and pole position $z_{1}$ of the corresponding
coupled-channel version of the potential, shown in Table~\ref{phys_char2.tab}.
The second one, $V^{\bar{K}N, \rm{Complex}}_{(a,a)}$ provides the same $I=0$ and
$I=1$ isospin $\bar{K}N$ scattering lengths as the full $\bar{K}N - \pi \Sigma$:
\begin{eqnarray}
&{}& a_{\bar{K}N, I=0}^{\rm one-pole} =  -1.60 + i \, 0.67 \,{\rm fm,} \quad
a_{\bar{K}N, I=0}^{\rm two-pole} =  -1.62 + i \, 0.78 \,{\rm fm,} \\
&{}& a_{\bar{K}N, I=1}^{\rm one-pole} =  -0.004 + i \, 0.57 \,{\rm fm,} \quad
a_{\bar{K}N, I=1}^{\rm two-pole} =  0.18 + i \, 0.68 \,{\rm fm.}
\end{eqnarray}

\subsection{Fixed center approximation}
\label{FCA.subsect}

So-called ``Fixed center approximation to Faddeev equations'' (FCA) introduced
in~\cite{Kd_KOR} is a variant of FSA or a two-center formula.
Fixed-scatterer approximation (FSA) or a two-center problem assumes, that
the scattering of a projectile particle takes place on two much heavier target
particles separated by a fixed distance. The motion of the heavy particles is
subsequently
taken into account by averaging the obtained projectile-target amplitude over
the bound state wave function of the target. Therefore, the approximation 
is well-known and works properly in atomic physics, were an electron is really
much lighter than a nucleon or an ion. Having in mind, that the antikaon mass is only
twice smaller than the mass of a nucleon, we can expect, that FSA hardly can be a
good approximation for the $K^- d$ scattering length calculation.

The FCA formula was obtained in~\cite{Kd_KOR} from Faddeev equations in a very
strange way. Proper derivation of a FSA formula starting from the same equations
was done much earlier in~\cite{peresypkin}, it can also be found in~\cite{Deloff_book}
together with several versions of the FSA formula.
Fixed scatterer approximation for the calculation of $a_{K^- d}$ scattering length
using separable potentials was used in~\cite{Kd_two-center}.

A novelty of the FCA formula of~\cite{Kd_KOR} is introduction of
``isospin breaking terms'', which, according to the authors, come from $\bar{K}^0 n$
two-body particle channel introduced in addition to $K^- p$. However,
the inclusion of the $\bar{K}^0 n$ channel is questionable, since, as it
was already mentioned in Section~\ref{AGS_eq.sect}, all terms, connected with
this subsystem, automatically drop out from the Faddeev system of equations after
antisymmetrization.

Necessity to go beyond FCA formula for the $K^- d$ system
was clearly demonstrated in~\cite{Kd_BFMS},
were the unstable character of the FCA results was pointed out. However, the
formula is still being used, for example in~\cite{ruzecky}, that is why we
decided to check the approximation. We used the same two-body input as in
AGS equations in order to make the comparison as adequate as possible.

First of all, we used the scattering lengths provided by our coupled-channel
$\bar{K}N - \pi \Sigma$ potentials and the deuteron wave function corresponding
to our TSA-B $NN$ potential in FCA formula~Eq.(24) from~\cite{Kd_KOR}.
Secondly, all $\bar{K}^0 n$ parts were removed from the formula because
they do not enter AGS equations. Finally, we took into account the fact,
that the FCA formula was obtained for a local $\bar{K}N$ potential, while separable
$\bar{K}N - \pi \Sigma$ potentials were used in our Faddeev equations.
The corresponding changes in the FCA formula were made 
\footnote{
To remove the difference we used $G(R)$ function defined by Eq.(25)
in~\cite{Kd_two-center} instead of $1 \slash R$ function, which is an
approximation of the $G(R)$, in the FCA formula. The range parameter
$\beta^{\bar{K}N}$, entering the $G(R)$ function, was also taken from our
coupled-channel $\bar{K}N - \pi \Sigma$ potential. In fact, the
replacement $1 \slash R$ by $G(R)$ changed the results negligibly.}.
Therefore, the two-body input for the FCA formula was equivalent to the input for
the AGS calculation.

\section{Results and discussion}
\label{Results.sect}

\begin{figure}
\centering
\includegraphics[width=0.35\textwidth, angle=-90]{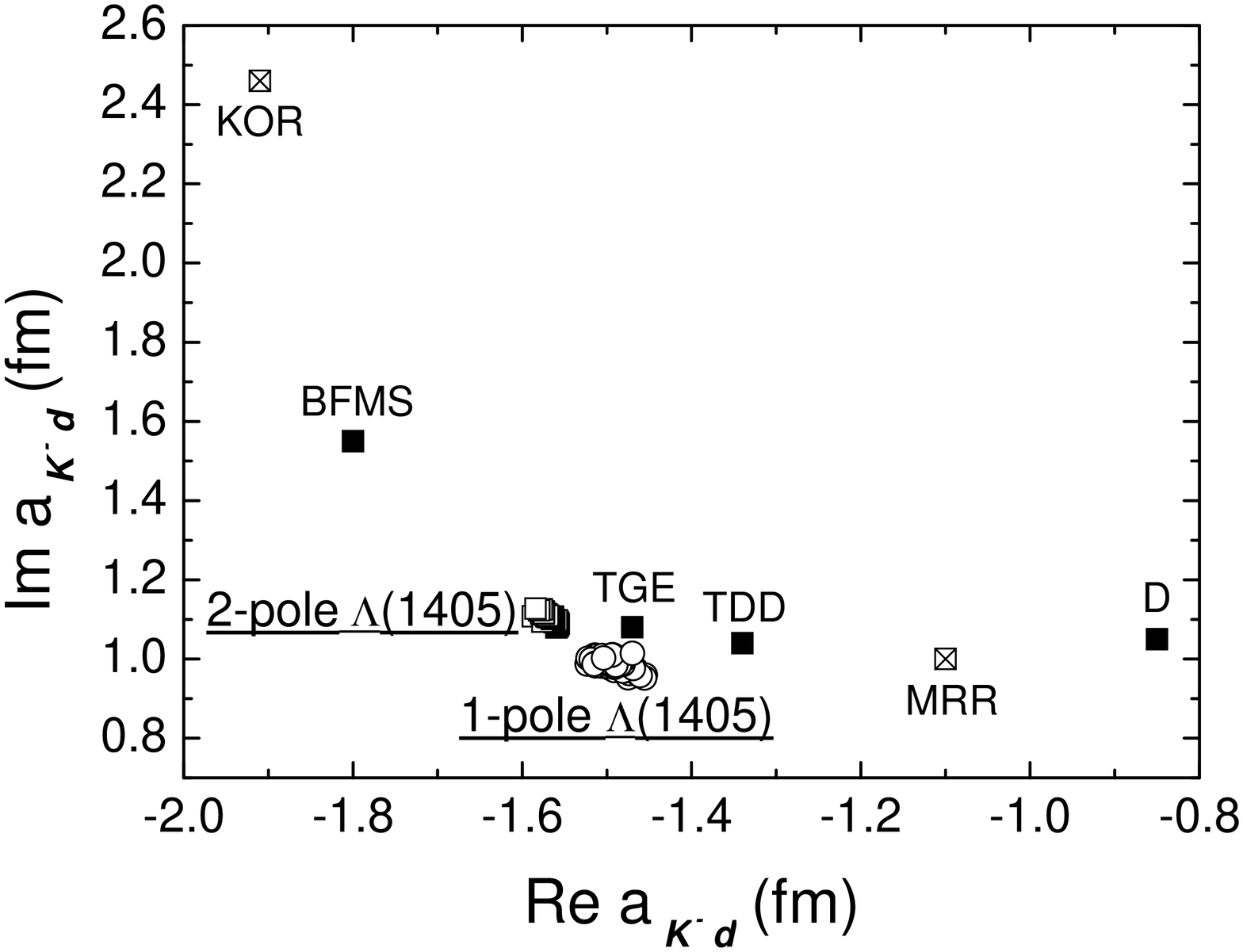}
\caption{The results of the full $K^- d$ scattering length calculations
using sets of one- (empty circles) and two-pole (empty squares) versions of
$\bar{K}N - \pi \Sigma$ potential. Previous Faddeev calculations:
BFMS~\cite{Kd_BFMS}, TGE~\cite{Kd_TGE}, TDD~\cite{Kd_TDD}, D~\cite{Kd_Deloff}
(filled squares), and FCA: KOR~\cite{Kd_KOR}, MRR~\cite{ruzecky}
(crossed squares) results are also shown.
\label{Kd_others}}
\end{figure}

The results of the full coupled-channel calculations of the $ K^- d$ scattering
length using sets of one- (empty circles) and two-pole (empty squares) versions
of the coupled-channel $\bar{K}N - \pi \Sigma$ potentials are shown in
Fig.\ref{Kd_others}. The calculations were performed with $V_{NN}^{TSA-B}$ and
exact optical $V_{\Sigma N}^{\rm Sdep, Opt}$. The $K^- p$ scattering
length values obtained with the two representative $\bar{K}N - \pi \Sigma$ potentials
are:
\begin{eqnarray}
a_{K^- d}^{\rm one-pole} &=& -1.49 + i \, 0.98 \, {\rm fm,} \\
a_{K^- d}^{\rm two-pole} &=& -1.57 + i \, 1.11 \, {\rm fm.}
\end{eqnarray}
Results of previous Faddeev calculations of the same system (filled squares) together
with two FCA results (crossed squares) are also shown in the figure.

It is seen, that while two-body data do not allow to distinguish between one- and
two-pole versions of ``the main'' $\bar{K}N - \pi \Sigma$ interaction, the three-body
sets of results differ sufficiently for such a task. Therefore, in principle, it would
be possible to favor one version of the $\bar{K}N - \pi \Sigma$ potential
by comparing with an experimental result. However, direct measurement of 
$K^- d$ scattering length is impossible. Moreover, it is not absolutely clear,
whether the difference between the two sets of the $a_{K^- d}$ results is much more than
theoretical uncertainties, caused mainly by the uncertainties of the $\bar{K}N$
model of interaction. In any case, a calculation of $1s$ level shift
$\Delta E_{\rm{deu}, 1s}$ and width $\Gamma_{{\rm deu}, 1s}$
of kaonic deuterium atom, corresponding to the obtained $a_{K^- d}$ values,
is necessary for comparing with experimentally measurable values. The parameters
of kaonic deuterium 
were being measured by SIDDHARTA experiment, unfortunately, without any results. 
Due to this our next step will be making predictions for $\Delta E_{\rm{deu}, 1s}$
and $\Gamma_{{\rm deu}, 1s}$ observables.

As is seen in Fig~\ref{Kd_others}, our $a_{K^- d}$ results are close to the other
$K^- d$ scattering lengths obtained in~\cite{Kd_TGE} and~\cite{Kd_TDD} within
coupled-channel Faddeev approach. The result of~\cite{Kd_Deloff}, obtained by a
one-channel Faddeev calculation with a zero-range one-channel $\bar{K}N$ potential,
has much smaller absolute value of the real part than all other $a_{K^- d}$. On
the contrary, the
authors of~\cite{Kd_BFMS}, who performed Faddeev calculations using $NN$ interaction
with $d$-wave component, obtained $K^- d$ scattering length 
with quite larger absolute values of both real and imaginary parts.

The $a_{K^- d}$ value of~\cite{Kd_KOR}, significantly different from all others, was
calculated using the FCA formula, obtained in the same paper and already discussed in
Section~\ref{FCA.subsect}. We chose the result calculated in
isospin basis for the
comparison. One more paper, where FCA formula was used, is~\cite{ruzecky}, where,
however, the result was obtained by simply applying of two approximate formulae.
The second one is the corrected Deser formula, used for calculation of the $\bar{K}N$
scattering lengths, entering the FCA. One of the representative $a_{K^- d}$ values
from~\cite{ruzecky}, having the largest possible imaginary part, is shown in
Fig.\ref{Kd_others}.

\begin{figure}
\centering
\includegraphics[width=0.35\textwidth, angle=-90]{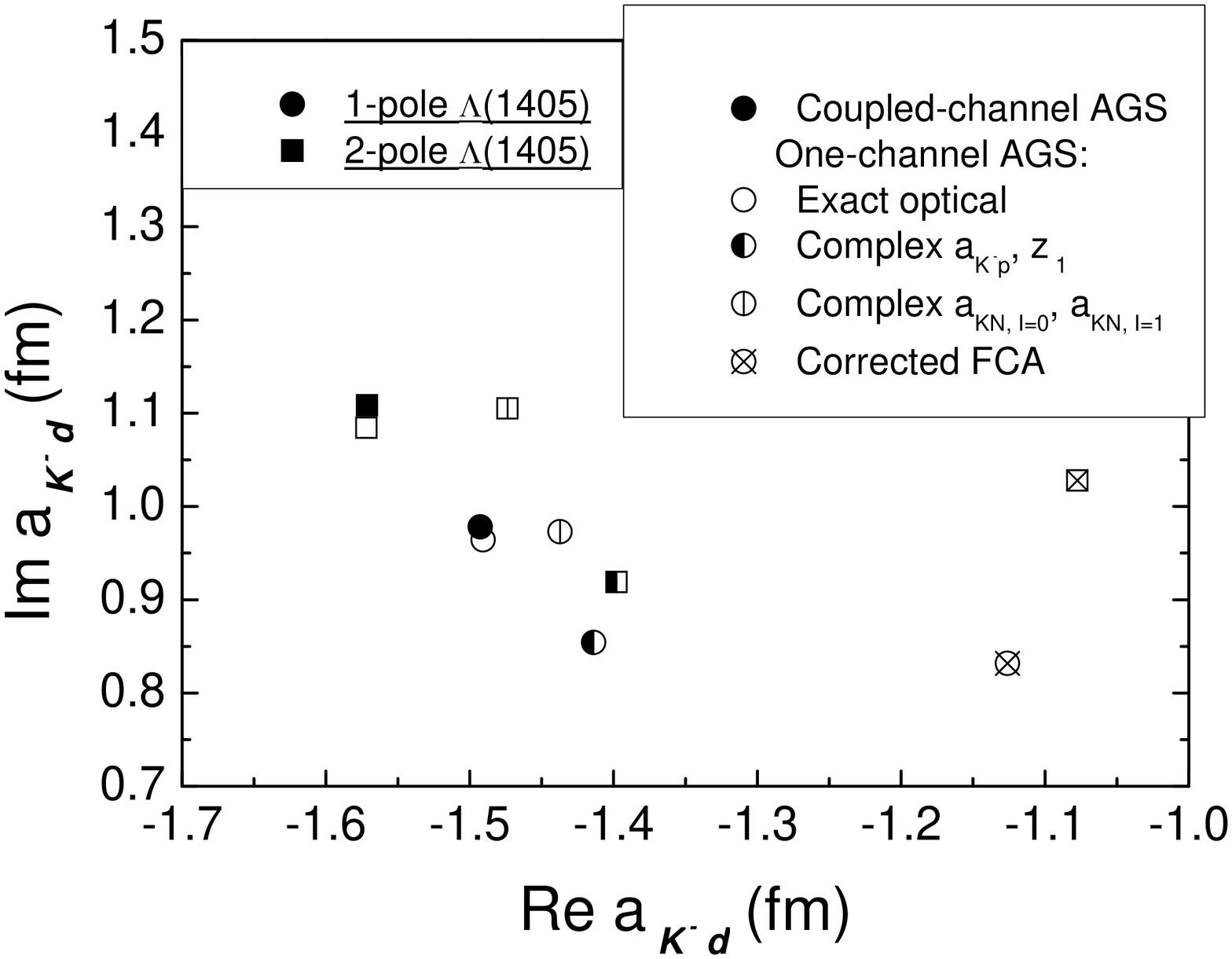}
\caption{Comparison of the full results of the $a_{K^- d}$ calculations
with representative one- (circles) and two-pole (squares)
versions of $\bar{K}N - \pi \Sigma$ potential with approximate results.
Values obtained with coupled-channel  AGS equations (filled symbol),
one-channel AGS with exact optical $\bar{K}N$ potential $V^{\bar{K}N}, \rm{opt}$
(empty symbol), one-channel AGS with complex $V^{\bar{K}N}, \rm{complex}_{(a,z)}$
(half-empty symbol) and $V^{\bar{K}N}, \rm{complex}_{(a,a)}$ (vertically crossed symbol)
results are shown. Results of corrected FCA formula using (crossed symbol) are also
shown.
\label{Kd_approx}}
\end{figure}

It is hard to compare all $a_{K^- d}$ results, because the methods of 
treatment of the three-body
problem and two-body inputs are different in the mentioned works.
In order to investigate separate effects of several approximations we performed
approximate calculations, as described in Section~\ref{approximate.sect}. The
obtained one-channel AGS $a_{K^- d}$ values (Section~\ref{one-channel.subsect})
together with FCA (Section~\ref{FCA.subsect}) and the representative coupled-channel AGS
results of the $K^- d$ scattering length calculations are shown in Fig.~\ref{Kd_approx}
for one- and two-pole versions of $\bar{K}N - \pi \Sigma$ interaction. It is important,
that all results in the figure were obtained with equivalent two-body
input, including the neglect of isospin-breaking parts in the original FCA formula.

It is seen from Fig.\ref{Kd_approx}, that all approximations are more
accurate for the one-pole version of the $\bar{K}N$ interaction than for the two-pole
variant.
The one-channel AGS calculation with exact optical $\bar{K}N$ potential (empty symbols),
giving exactly the same $\bar{K}N - \bar{K}N$ amplitude as the corresponding
coupled-channel potential, turns out to be the best approximation. The result, obtained
with the simple complex potential $V^{\bar{K}N, \rm{complex}}_{(a,a)}$
(vertically crossed symbols), reproducing $I=0$ and $I=1$ $\bar{K}N$
scattering lengths, underestimates the absolute
value of the real part of the $a_{K^- d}$, especially
for the two-pole version of $\bar{K}N$ interaction, but have rather accurate
imaginary part of the $K^- d$ scattering length. Another one-channel AGS calculation with
simple complex potential $V^{\bar{K}N, \rm{complex}}_{(a,z)}$, reproducing $K^- p$
scattering length and
pole position $z_{1}$, gives rather inaccurate result (half-empty symbols) as
compared with the coupled-channel AGS values (filled symbols).

The scattering length $a_{K^- d}$ of~\cite{Kd_Deloff} was obtained from one-channel
Faddeev equations with a complex potential. However, the underestimation of 
the absolute values of its real part
in comparison to other Faddeev calculations is so large, that it cannot be explained
by the method only. The most likely reason of the difference lyes in the
properties of the $\bar{K}N$ potential used in~\cite{Kd_Deloff}. First of all,
the potential has a very large  mass of the $K^- p$ quasibound state ($1439$ MeV),
which, therefore, is situated above the $K^- p$ threshold. In addition, the width of the
state ($127$ MeV) is much larger than the PDG value ($50$ MeV) as well as the width of
our $z_1$ pole ($72$ MeV).

It is hard to understand the results obtained in~\cite{Kd_BFMS}. While all formulae
are written for many-channel Faddeev equations, the most of the dependences and
even ``the best'' $a_{K^- d}$ value were obtained within a one-channel Faddeev
calculation including $\bar{K}N$ and $NN$ interactions only. Since the elastic
part of the coupled-channel $\bar{K}N$ $T$-matrix
was used, the result is equivalent to a one-channel Faddeev calculation
with an exact optical potential. But even the full coupled-channel calculation
was performed in~\cite{Kd_BFMS} with non-unitary $\bar{K}N$ $T$-matrices OSA and
OS1, since channels with $\eta$-mesons, entering the two-body $T$, were omitted in
the three-body equations. It is not clear, why the one-channel calculations
of~\cite{Kd_BFMS} give so large difference in imaginary parts of $a_{K^- d}$ obtained
in isospin and particle basis and with and without $d$-wave in $NN$. The result of a
calculation, in principle, should not depend on the chosen basis, in addition, the very
recent results of $K^- d$ scattering calculations~\cite{Kd_Janos} demonstrated, that 
inclusion of physical masses into Faddeev equations change $a_{K^- d}$ by
several percents only.

The results of using the FCA formula without isospin-breaking effects (crossed symbols)
stays far away from the full calculation, as is seen in Fig.\ref{Kd_approx}. While
errors for the imaginary part are not so large, the module of the real part is
underestimated by about $30 \%$. Therefore, our calculations show, that FCA is a poor
approximation for $K^- d$ scattering length calculation, and the accuracy is lower
for the two-pole $\bar{K}N$ model of interaction (the most of chirally-based
models of $\bar{K}N$ interaction have two-pole structure).
Even the original FCA formula does not give such a large $K^- d$ scattering length as 
$a_{K^- d}$ from~\cite{Kd_KOR}, which, therefore, is caused by too large
input $\bar{K}N$ scattering lengths. As for the values of~\cite{ruzecky},
they suffer from cumulative errors of two approximations and using of DEAR
results on kaonic hydrogen characteristics. As was shown
in~\cite{ourPRC_isobreak}, the error of the corrected Deser formula makes
about $10 \%$, while the problems with DEAR experimental data were also
demonstrated in the paper and in other theoretical works.

\begin{figure}
\centering
\includegraphics[width=0.35\textwidth, angle=-90]{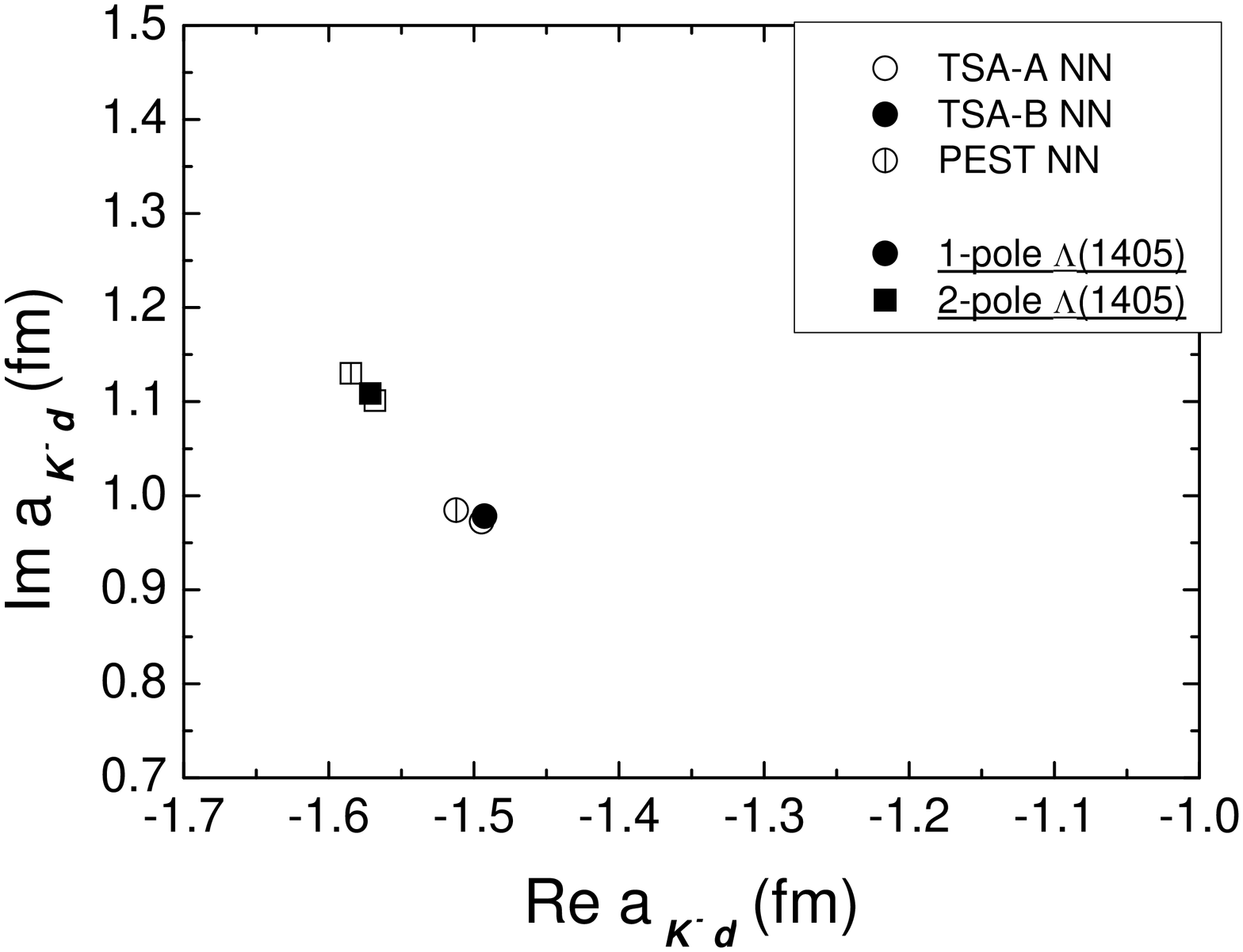}
\caption{Dependence of the full $K^- d$ scattering length results
on $NN$ interaction: $a_{K^- d}$ obtained with with TSA-A (empty symbol),
TSA-B (filled symbol), and PEST (vertically crossed symbol) $NN$ potentials.
One- (circles) and two-pole (squares) representative versions of
$\bar{K}N - \pi \Sigma$ interaction were used.
\label{Kd_NNdep}}
\end{figure}

We investigated dependence of the full coupled-channel results on $NN$ and
$\Sigma N (-\Lambda N)$ interactions as well.
The dependence of $a_{K^- d}$ on nucleon-nucleon interaction
is demonstrated in Fig~\ref{Kd_NNdep}, were the results obtained with TSA-A, TSA-B,
and PEST $V^{NN}$, are shown. The representative sets of one- and two-pole
$\bar{K}N - \pi \Sigma$ potentials were used together with exact optical $V^{\rm Sdep, Opt}$
$\Sigma N (-\Lambda N)$ model of interaction. We see from the Figure, that the difference
is very small even for the potentials with and without repulsion
at short distances (TSA and PEST, correspondingly). Therefore, the $s$-wave $NN$ interaction,
which is used in the present calculation, plays minor role in the calculation. Most likely,
it is caused by relative weakness of the $NN$ interaction as compared to $\bar{K}N$  from the
viewpoint of a much deeper quasibound state in the latter system ($E_{\bar{K}N} \approx - 23$
MeV for our potentials) than the deuteron bound state ($E_{\rm deu} \approx -2$ MeV).
We do not expect much larger effect from higher partial waves in $NN$ as well.

\begin{figure}
\centering
\includegraphics[width=0.35\textwidth, angle=-90]{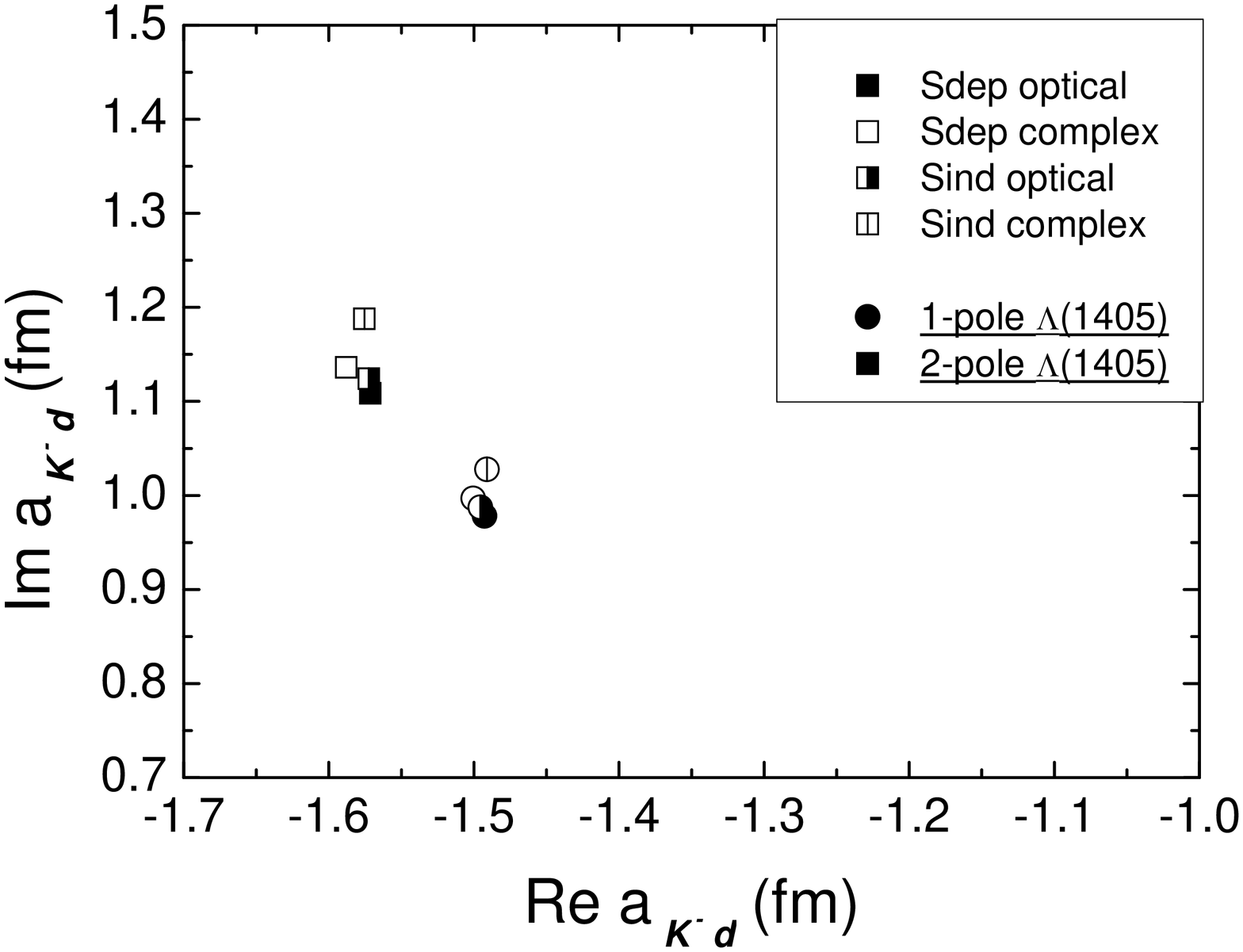}
\caption{Dependence of the full $K^- d$ scattering length results
on $\Sigma N - (\Lambda N)$ interaction. The $a_{K^- d}$ values obtained with
exact optical $V^{\rm Sdep, opt}$ (filled square), simple complex
$V^{\rm Sdep, complex}$ (empty square), exact optical $V^{\rm Sind, opt}$
(half-empty square), and simple complex $V^{\rm Sind, complex}$
(vertically crossed square) are shown. One- (circles) and two-pole
(squares) representative versions of $\bar{K}N - \pi \Sigma$ interaction were used.
\label{Kd_SigmaNdep}}
\end{figure}

We also looked at the dependence of $a_{K^- d}$ on $\Sigma N (-\Lambda N)$ interaction.
The $K^- d$ scattering lengths obtained with the exact optical and the simple complex
versions of the spin dependent
$V^{\rm Sdep}$ and spin independent $V^{\rm Sind}$ potentials
are shown in Fig.~\ref{Kd_SigmaNdep}.
The representative sets of one- and two-pole $\bar{K}N - \pi \Sigma$
potentials were used together with TSA-B $NN$ potential.
The results of the two versions of $\Sigma N (-\Lambda N)$ potential 
$V^{\rm Sdep}$ and $V^{\rm Sind}$ in exact optical
form are very close, while their simple complex versions are slightly different.
However, the largest error does not exceed $3 \%$, therefore, the dependence of
the $a_{K^- p}$ on $\Sigma N - (\Lambda N)$ interaction is also weak.

\section{Conclusions}
\label{conclusions.sect}

To conclude, we performed calculations of the $K^- d$ scattering length using
newly obtained coupled-channel $\bar{K}N - \pi \Sigma$ potentials
with one- and two-pole versions of the $\Lambda(1405)$ resonance. 
Faddeev-type AGS equations were used for description of $\bar{K}NN - \pi \Sigma N$
system. We also constructed new coupled-channel
$\Sigma N -\Lambda N$ potentials together with its exact optical and simple complex
$\Sigma N (-\Lambda N)$ versions.
Different models of the $NN$ interaction: TSA-A, TSA-B, and PEST were used.
All two-body interactions are described by $s$-wave separable potentials.
We investigated dependence of $a_{K^- d}$ on $NN$ and $\Sigma N (-\Lambda N)$
interaction and found, that both dependences are weak.

We found, that the two sets of the results, obtained with one- and two-pole
models of $\Lambda(1405)$ resonance, are clearly separated one from another,
in principle, allowing to give preference to one of the $\bar{K}N - \pi \Sigma$
interaction models. However, the question, whether theoretical uncertainties
are not of the same order as the differences between the two obtained sets of
$a_{K^- d}$, remains open.
In any case, it is necessary to calculate level shifts and widths of kaonic deuterium
atom, corresponding to the obtained $K^- d$ scattering lengths, which can be
measured, say, by SIDDHARTA-2 experiment. It is assumed to be done in a next paper.

Among approximate results, the one-channel AGS calculation with exact optical
$\bar{K}N (-\pi \Sigma)$ potential gives the best approximation
to the full coupled-channel result. On the contrary, FCA was shown to be the
least accurate approximation, especially in reproduction of the real part
of the $K^- d$ scattering length. All approximations are less accurate for
the two-pole model of $\bar{K}N - \pi \Sigma$ interaction.

%%%%%%%%%%%%%%%%%%%%%%%%%%%%%%%%%%%%%%%%%%%%%%%%%%%%%%%%%%%%%%%%%%%%%%%%%%
%%%%%%%%%%%%%%%%%%%%%%%%%%%%%%%%%%%%%%%%%%%%%%%%%%%%%%%%%%%%%%%%%%%%%%%%%%

%%%%%%%%%%%%%%%%%%%%%%%%%%%%%%%%%%%%%%%%%%%%%%%%%%%%%%%%%%%%%%%%%%%%%%%%%%
%%%%%%%%%%%%%%%%%%%%%%%%%%%%%%%%%%%%%%%%%%%%%%%%%%%%%%%%%%%%%%%%%%%%%%%%%%

%%%%%%%%%%%%%%%%%%%%%%%%%%%%%%%%%%%%%%%%%%%%%%%%%%%%%%%%%%%%%%%%%%%%%%%%%%
%%%%%%%%%%%%%%%%%%%%%%%%%%%%%%%%%%%%%%%%%%%%%%%%%%%%%%%%%%%%%%%%%%%%%%%%%%

%%%%%%%%%%%%%%%%%%%%%%%%%%%%%%%%%%%%%%%%%%%%%%%%%%%%%%%%%%%%%%%%%%%%%%%%%%
%%%%%%%%%%%%%%%%%%%%%%%%%%%%%%%%%%%%%%%%%%%%%%%%%%%%%%%%%%%%%%%%%%%%%%%%%%

%%%%%%%%%%%%%%%%%%%%%%%%%%%%%%%%%%%%%%%%%%%%%%%%%%%%%%%%%%%%%%%%%%%%%%%%%%
%%%%%%%%%%%%%%%%%%%%%%%%%%%%%%%%%%%%%%%%%%%%%%%%%%%%%%%%%%%%%%%%%%%%%%%%%%

%%%%%%%%%%%%%%%%%%%%%%%%%%%%%%%%%%%%%%%%%%%%%%%%%%%%%%%%%%%%%%%%%%%%%%%%%%
%%%%%%%%%%%%%%%%%%%%%%%%%%%%%%%%%%%%%%%%%%%%%%%%%%%%%%%%%%%%%%%%%%%%%%%%%%

%%%%%%%%%%%%%%%%%%%%%%%%%%%%%%%%%%%%%%%%%%%%%%%%%%%%%%%%%%%%%%%%%%%%%%%%%%
%%%%%%%%%%%%%%%%%%%%%%%%%%%%%%%%%%%%%%%%%%%%%%%%%%%%%%%%%%%%%%%%%%%%%%%%%%

%%%%%%%%%%%%%%%%%%%%%%%%%%%%%%%%%%%%%%%%%%%%%%%%%%%%%%%%%%%%%%%%%%%%%%%%%%
%%%%%%%%%%%%%%%%%%%%%%%%%%%%%%%%%%%%%%%%%%%%%%%%%%%%%%%%%%%%%%%%%%%%%%%%%%

%%%%%%%%%%%%%%%%%%%%%%%%%%%%%%%%%%%%%%%%%%%%%%%%%%%%%%%%%%%%%%%%%%%%%%%%%%
%%%%%%%%%%%%%%%%%%%%%%%%%%%%%%%%%%%%%%%%%%%%%%%%%%%%%%%%%%%%%%%%%%%%%%%%%%

%%%%%%%%%%%%%%%%%%%%%%%%%%%%%%%%%%%%%%%%%%%%%%%%%%%%%%%%%%%%%%%%%%%%%%%%%%
%%%%%%%%%%%%%%%%%%%%%%%%%%%%%%%%%%%%%%%%%%%%%%%%%%%%%%%%%%%%%%%%%%%%%%%%%%

%%%%%%%%%%%%%%%%%%%%%%%%%%%%%%%%%%%%%%%%%%%%%%%%%%%%%%%%%%%%%%%%%%%%%%%%%%
%%%%%%%%%%%%%%%%%%%%%%%%%%%%%%%%%%%%%%%%%%%%%%%%%%%%%%%%%%%%%%%%%%%%%%%%%%

%%%%%%%%%%%%%%%%%%%%%%%%%%%%%%%%%%%%%%%%%%%%%%%%%%%%%%%%%%%%%%%%%%%%%%%%%%
%%%%%%%%%%%%%%%%%%%%%%%%%%%%%%%%%%%%%%%%%%%%%%%%%%%%%%%%%%%%%%%%%%%%%%%%%%

%%%%%%%%%%%%%%%%%%%%%%%%%%%%%%%%%%%%%%%%%%%%%%%%%%%%%%%%%%%%%%%%%%%%%%%%%%
%%%%%%%%%%%%%%%%%%%%%%%%%%%%%%%%%%%%%%%%%%%%%%%%%%%%%%%%%%%%%%%%%%%%%%%%%%

\vspace{5mm}

\noindent
{\bf Acknowledgments.}
The author is thankful to J. R\'evai for many fruitful discussions and to
J. Haidenbauer for his comments concerning $\Sigma N - \Lambda N$
interaction. The work was supported by the Czech GA AVCR grant KJB100480801.


\begin{thebibliography}{99}

\bibitem{ECT2009} Mini-Proceedings ECT* Workshop ''Hadronic Atoms and Kaonic Nuclei'',
(ECT*, Trento, Italy, October 12--16, 2009),
Eds. C. Curceanu and J. Marton; nucl-ex/1003.2328.

\bibitem{Faddeev} L.D. Faddeev, Soviet Phys. JETP 12, 1014 (1961);
Mathematical aspects of the three-body problem in quantum scattering theory,
Steklov Math. Institute 69 (1963).

\bibitem{ourPRC_isobreak} J. R\'evai, N.V. Shevchenko,
Phys. Rev. C 79, 035202 (2009).

\bibitem{SIDDHARTA} C. Curceanu {\it et al}., Eur. Phys. J. A 31, 537 (2007).

\bibitem{AGS} E.O. Alt, P. Grassberger, W. Sandhas, Nucl. Phys. B 2, 167 (1967).

\bibitem{our_KNN_PRL} N.V. Shevchenko, A. Gal, J. Mare\v{s}, Phys. Rev. Lett. 98,
082301 (2007).

\bibitem{our_KNN_PRC} N.V. Shevchenko, A. Gal, J. Mare\v{s}, J. R\'evai,
Phys. Rev. C 76, 044004 (2007).

\bibitem{VB} V.B. Belyaev, {\textit{Lectures on the Theory of Few-Body Systems}},
(Springer Verlag, 1990).

\bibitem{log_sing} F. Sohre and H. Ziegelman, Phys. Lett. B 34, 579 (1971).

\bibitem{gammaKp1} D.N. Tovee {\it et al.}, Nucl. Phys. B 33, 493 (1971).

\bibitem{gammaKp2} R.J.~Nowak {\it et al.}, Nucl. Phys. B 139, 61 (1978).

\bibitem{Kp2exp} M. Sakitt {\it et al.}, Phys. Rev. 139, B719 (1965).

\bibitem{Kp3exp} J.K. Kim, Phys. Rev. Lett. 14, 29 (1965);
Columbia University Report, Nevis, 149 (1966); Phys. Rev. Lett. 19, 1074 (1967).

\bibitem{Kp4exp} W. Kittel, G. Otter, and I. Wacek, Phys. Lett. 21, 349 (1966).

\bibitem{Kp5exp} J. Ciborowski {\it et al.}, J. Phys. G 8, 13 (1982).

\bibitem{Kp6exp} D. Evans {\it et al.}, J. Phys. G 9, 885 (1983).

\bibitem{Kp1exp} W.E. Humphrey, R.R. Ross, Phys. Rev. 127, 1305 (1962).

\bibitem{KEK1s} M. Iwasaki {\it et al.}, Phys. Rev. Lett. 78, 3067 (1997);
T.M.~Ito {\it et al.}, Phys. Rev. C 58, 2366 (1998).

\bibitem{DEAR1s} G. Beer {\it et al.}, Phys. Rev. Lett. 94, 212302 (2005).

\bibitem{SIDDHARTA1s} M. Bazzi {\it et al.}, nucl-ex/1105.3090.

\bibitem{Borasoy_aKp} B. Borasoy, U.-G. Mei{\ss}ner, R. Ni{\ss}ler,
Phys. Rev. C 74, 055201 (2006).

\bibitem{PDG} K. Nakamura {\it et al.} (Particle Data Group), J. Phys. G 37, 075021 (2010).

\bibitem{CLAS} R. Schumacher (for the CLAS Collaboration), AIP Conf. Proc. 1257, 100 (2010).

\bibitem{DolesNN} P. Doleschall, {\it private communication}.

\bibitem{ArgonneV18} R. B. Wiringa, V. G. J. Stoks, and R. Schiavilla,
Phys. Rev. C 51, 38 (1995).

\bibitem{NNpot} H. Zankel, W. Plessas, J. Haidenbauer, Phys. Rev. C 28, 538 (1983).

% Sigma N - Lambda N advanced potentials:
\bibitem{SigmaNth1} J. Haidenbauer, U.-G. Meissner, Phys. Rev. C 72, 044005 (2005).

\bibitem{SigmaNth2} Th. A. Rijken, Y. Yamamoto, Phys. Rev. C 73, 044008 (2006).

% Sigma N - Lambda N exp. data:
\bibitem{SigmaN1} G. Alexander, U. Karshon, A. Shapira, G. Yekutieli,
R. Engelmann, H. Filthuth, and W. Lughofer, Phys. Rev. 173, 1452 (1968).

\bibitem{SigmaN2} B. Sechi-Zorn, B. Kehoe, J. Twitty, and R. A. Burnstein,
Phys. Rev. 175, 1735 (1968).

\bibitem{SigmaN3} F. Eisele, H. Filthuth, W. F\"{o}lisch, V. Hepp, E. Leitner,
and G. Zech, Phys. Lett. B 37, 204 (1971).

\bibitem{SigmaN4} R. Engelmann, H. Filthuth, V. Hepp, and E. Kluge,
Phys. Lett. 21, 587 (1966).

\bibitem{SigmaN5} V. Hepp and M. Schleich, Z. Phys. 214, 71 (1968).

\bibitem{Kd_KOR} S.S. Kamalov, E. Oset, A. Ramos, Nucl. Phys. A 690, 494 (2001).

\bibitem{peresypkin} V.V. Peresypkin, Ukr. Fiz. Zh. 23, 1256 (1978).

\bibitem{Deloff_book} A. Deloff, {\textit{Fundamentals in hadronic atom theory}},
(World scientific, 2003).

\bibitem{Kd_two-center} R. C. Barrett, A. Deloff, Phys. Rev. C 60, 025201 (1999).

\bibitem{Kd_BFMS} A. Bahaoui, C. Fayard, T. Mizutani, B. Saghai, Phys. Rev. C 68, 064001 (2003).

% MRR point is one of the representative points from:
\bibitem{ruzecky} U.-G. Meissner, U. Raha, A. Rusetsky, Eur. Phys. J. C 47, 473 (2006).

\bibitem{Kd_TGE} G. Toker, A. Gal, J.M. Eisenberg, Nucl. Phys. A 362, 405 (1981).

\bibitem{Kd_TDD} M. Torres, R.H. Dalitz, A. Deloff, Phys. Lett. B 174, 213 (1986).

\bibitem{Kd_Deloff} A. Deloff, Phys. Rev. C 61, 024004 (2000).

\bibitem{Kd_Janos} J. R\'evai, {\it to be published.}

\end{thebibliography}
\end{document}